\documentclass[letterpaper,journal]{IEEEtran} 
\usepackage{amsmath,amsfonts}
\usepackage{algorithmic}
\usepackage{algorithm}
\usepackage{array}
\usepackage[caption=false,font=normalsize,labelfont=sf,textfont=sf]{subfig}
\usepackage{textcomp}
\usepackage{stfloats}
\usepackage{url}
\usepackage{verbatim}
\usepackage{graphicx}
\usepackage{cite}
\usepackage{multirow}

\newcommand*{\defeq}{\mathrel{\vcenter{\baselineskip0.5ex \lineskiplimit0pt
                     \hbox{\scriptsize.}\hbox{\scriptsize.}}}%
                     =}

\begin{document}

\title{Resilience and Load Balancing in Fog Networks:\\ A Multi-Criteria Decision Analysis Approach}

\author{Maad Ebrahim, Abdelhakim Hafid,~\IEEEmembership{Member,~IEEE}
\thanks{The authors are with the NRL, Department of Computer Science and Operational Research, University of Montreal, Montreal, QC H3T-1J4, Canada (e-mail: maad.ebrahim@umontreal.ca; ahafid@iro.umontreal.ca).}
\thanks{Corresponding author: Maad Ebrahim (maad.ebrahim@umontreal.ca).}}

\maketitle

\begin{abstract}
The advent of Cloud Computing enabled the proliferation of IoT applications for smart environments. However, the distance of these resources makes them unsuitable for delay-sensitive applications. Hence, Fog Computing has emerged to provide such capabilities in proximity to end devices through distributed resources. These limited resources can collaborate to serve distributed IoT application workflows using the concept of stateless micro Fog service replicas, which provides resiliency and maintains service availability in the face of failures. Load balancing supports this collaboration by optimally assigning workloads to appropriate services, i.e., distributing the load among Fog nodes to fairly utilize compute and network resources and minimize execution delays. In this paper, we propose using ELECTRE, a Multi-Criteria Decision Analysis (MCDA) approach, to efficiently balance the load in Fog environments. We considered multiple objectives to make service selection decisions, including compute and network load information. We evaluate our approach in a realistic unbalanced topological setup with heterogeneous workload requirements. To the best of our knowledge, this is the first time ELECTRE-based methods are used to balance the load in Fog environments. Through simulations, we compared the performance of our proposed approach with traditional baseline methods that are commonly used in practice, namely random, Round-Robin, nearest node, and fastest service selection algorithms. In terms of the overall system performance, our approach outperforms these methods with up to 67\% improvement.
\end{abstract}


\section{Introduction}
\begin{figure}[!t]
    \centering
    \includegraphics[width=0.4\textwidth]{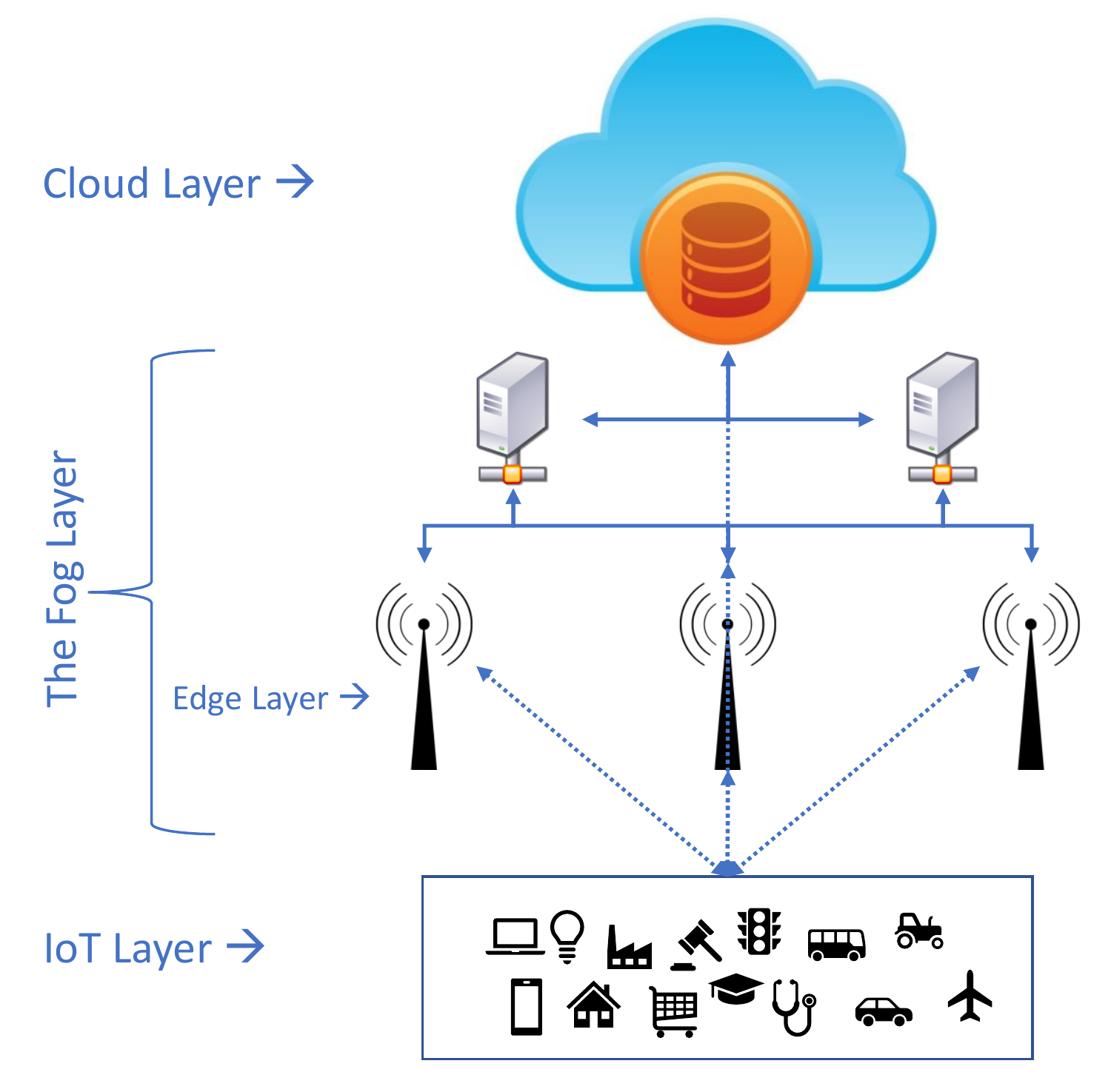}
    \caption{Cloud, Fog, and Edge Computing for IoT networks.\label{fig:FogEdge}}
\end{figure}

Fog Computing complements Cloud Computing to support delay-sensitive IoT applications and to support mobility, geo-distribution, and location awareness for these applications \cite{Survey2015}. It saves the network bandwidth by reducing the traffic between end devices and the Cloud. In addition, it increases the security and privacy of IoT applications by pre-processing and encrypting data closer to its source \cite{SecurityPrivacy}. Fog resources extend from the edge of the network to the Cloud, i.e., cloud-to-thing continuum, while Edge Computing limits these resources within one-hop distance from those things \cite{FogRef} (see Fig. \ref{fig:FogEdge}). Hence, the Fog is more complex than the Edge; indeed, the Edge can be viewed as a subcategory of the Fog as shown in Fig. \ref{fig:FogEdge}. Only modular applications, like workflow and bag-of-tasks \cite{contextaware}, allow for distributed Fog deployments as pipelined workflows \cite{Vertical}. In contrast, monolithic applications cannot be divided into multiple logical modules, and hence each shall run as a single module in a single computing entity.

\begin{figure*}[!t]
\centering
\subfloat[]{\includegraphics[trim=20 50 430 180, clip, width=0.48\textwidth]{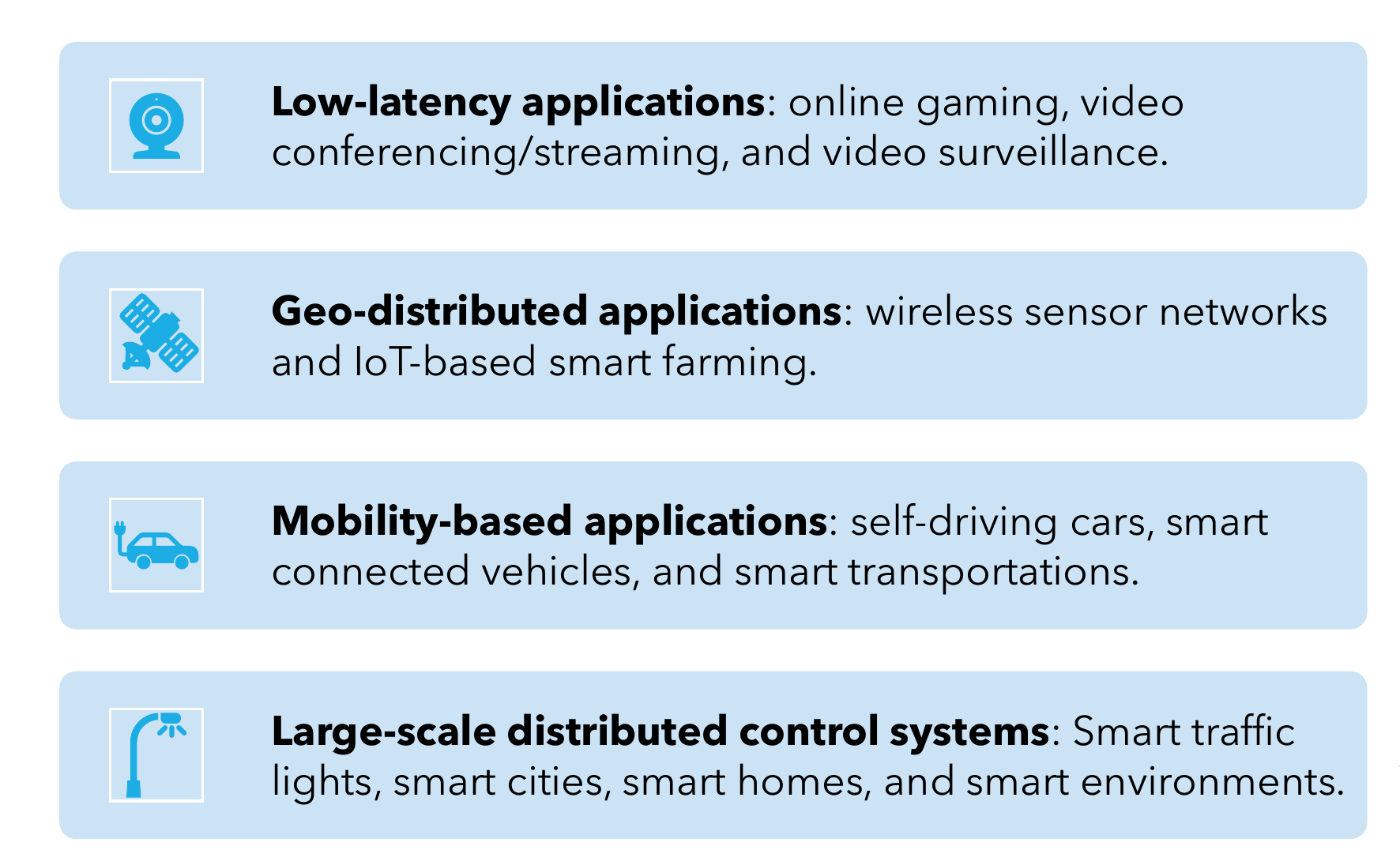}
\label{fig:AllApps}}
\hfil
\subfloat[]{\includegraphics[trim=520 50 20 180, clip, width=0.48\textwidth]{Figures/IoTApps.pdf}
\label{fig:VidApp}}
\caption{The benefits of distributed computing for IoT and mobile applications. (a) Edge and Fog-based IoT applications. (b) Fog-based video surveillance.}
\label{fig:IoTApps}
\end{figure*}

Many applications benefit from Fog and Edge Computing to support different types of IoT and mobile applications (see Fig. \ref{fig:IoTApps}\subref{fig:AllApps}). In Fig. \ref{fig:IoTApps}\subref{fig:VidApp}, we present a Fog-based video surveillance system, where the overall system performance is increased using pipelined application workflows. The first module, i.e., on the camera, performs simple subtractions between subsequent frames to prevent sending identical frames of stationary views. Edge and Fog modules perform object recognition and face detection, respectively; this allows to only send and process frames with humans and faces, respectively. Face recognition will run on the Cloud, where faces are matched against a privately-owned, or government-owned, database to track people in restricted areas, suspects, or criminals. Immediate feedback can be sent from each module in the application workflow to change frame resolution and video Bitrate in the sensing device when needed. For example, if a face is detected, the camera is notified to send higher resolution frames to increase the face recognition accuracy. However, the camera is notified to decrease its resolution and Bitrate to save resources when no objects are detected.

Estimating the volume of IoT workloads in advanced can help building Fog environments, from scratch or by extending existing infrastructures, to support IoT frameworks and their applications \cite{Dimensioning2}. However, it is almost impossible to accurately estimate the expected volume of IoT workloads, a problem that is imposed by the ever-growing number of IoT devices and their applications. Therefore, optimal resource management becomes critical to cope with these increasing demands, which is done through optimal resource provisioning and resource allocation schemes. In addition, optimally balancing IoT workload among Fog resources helps achieving optimal utilization of existing resources without requiring additional hardware resources. Although, optimal load balancing is not sufficient when the total available hardware resources can not provide enough computational power to serve all generated IoT workloads at a given point in time.

Time-dependent data rate fluctuations, uneven sensor distribution, and sensor mobility are usually the main reasons for dynamic workloads in realistic IoT networks. Hence, efficient management of the network’s scarce resources is essential to improve the performance of these systems. In mobile networks, for example, this requires understanding the behavior of network traffic \cite{MobilityPrediction}. These dynamic environment changes increase the complexity of efficient workload distribution across Fog nodes \cite{Survey2015}. This led researchers to oversimplify their experiments using many pre-assumptions, which makes their proposed solutions far from being applicable to realistic environments \cite{Comparison}. For example, Kashani et al. \cite{SystematicReview} proposed to use a three-layer hierarchical architecture, where the Cloud resides at the top, Fog nodes in the middle, and IoT devices at the bottom of the hierarchy. However, flat mesh-like architectures and semi-hierarchical architectures are more realistic, where Fog nodes can communicate with adjacent Fog nodes without the concept of layers \cite{Comparison}. This can be achieved by interconnecting Fog nodes with each other to allow offloading workloads and traffic among themselves, if needed, without necessarily going through the Cloud.

Load balancing becomes critical to achieving resource efficiency by avoiding bottlenecks, underload, and overload situations \cite{SystematicReview}. For this to work, the total available resources should be equal or greater than the requirements of all incoming requests at a given time \cite{LBEdge}. Hence, load balancing methods try to maximize resource utilization while minimizing the execution delay to satisfy the deadline requirements for delay-sensitive tasks. In addition, to distribute workloads among interconnected Fog nodes without migrating data and services, these nodes must have redundant service replicas of the requested modules \cite{loadDistribution}. 

Using redundant micro services for Fog-based applications is essential, especially for resource-hungry real-time IoT applications that are used globally, such as trending online games, Internet of Vehicles (IoV), and health monitoring systems. Balancing the load of such heavy applications is critical, and avoiding migration while serving the global population can be only achieved through service replication. These Fog-based micro services can be deployed globally as background services, and only loaded when triggered, i.e., when requested by an IoT application. Hence, the compute resources of those Fog nodes will only be consumed while serving these workloads. If workloads of some applications are less frequent in a specific region, their corresponding micro-services that are deployed in the Fog nodes in that region will be idle most of the time, hence saving their compute resources.

Considering redundant micro Fog services with resource-demanding real-time applications saves compute and network resources from the overhead of service and data migration. Migration overhead is a heavy burden in Fog networks, especially the overhead of transmission, allocation, and security of the migrated services and their associated data \cite{migrationCost}. In addition, redundancy allows performing load balancing through Fog service selection decisions, where a Fog service in a given Fog node is selected to serve a given request. Hence, an up-to-date directory of available services in every available Fog node must be presented for the node that performs load balancing. Such directory represents a virtual sub-network, i.e., local view of the network, for every workload, where multiple Fog nodes host the required service. Hence, the goal of the load balancing algorithm is to select one of these Fog nodes to serve this workload.

Service selection decisions depend on whether Fog modules need cached data to process IoT requests, a concept referred to as the state of the flow of requests \cite{SDLB}. For stateless requests, where cached data is not needed, incoming requests can be offloaded to other service replicas without migration \cite{SDLB}. However, migration is needed for stateful, i.e., not stateless, requests during the offloading process. Hence, to be able to use service selection decisions to balance the load between redundant micro Fog services while avoiding migration, workloads/requests must be assumed stateless, i.e., serverless. All these assumptions are common in Fog environments, in fact, considering stateless requests with redundant micro services provides resilience for the system by maintaining service availability in case of the failure of one or more Fog nodes in the system \cite{serverless}.

Service selection decisions range from sophisticated load balancing algorithms to simple random node selection \cite{DistributedLB}. Round-Robin (RR), nearest node, and fastest service selection algorithms are other simple service selection methods. Using the nearest node minimizes communication delay and energy consumption in the network while using the fastest service increases resource utilization and minimizes task processing time. But, smarter algorithms simultaneously include workload requirements, resource capabilities of computing nodes and network links, and their current load information. Search-based optimization algorithms can be used to balance the load in such fully observable environments, where the network load information is accessible to the load balancing algorithm.

Even though search-based algorithms may require higher processing power than simple traditional approaches, they can provide optimal, or near-optimal, results that can be used as a baseline for other algorithms. Such algorithms can be deployed in resource-rich network controllers, like SDN Controllers, to provide service selection decisions for every generated workload in the network. Therefore, we propose, in this paper, a multi-objective search-based optimization approach using an outranking Multi-Criteria Decision Analysis (MCDA) algorithm called ELECTRE \cite{ELECTRE}. Our approach outperformed traditional service selection algorithms, i.e., random, RR, nearest node, and fastest service selection algorithms, to improve the overall system performance. These traditional methods are widely considered for practical Fog deployments because of their simplicity while they are wrongly assumed to work well in arbitrary Fog architectures.

We considered using an MCDA-based approach to balance the load in Fog networks as it recently showed good performances in similar tasks. For instance, these methods achieved satisfactory performance in Mobile Crowd Computing systems by making optimal resource selection decisions \cite{CrowdComputing}. They were also used to solve the service selection problem in Cloud Computing environments \cite{CloudSelection}. In addition, these algorithms were used to solve the service placement problem in Fog Computing systems \cite{MCDA}. However, this is the first time an MCDA-based method is used to balance the load in Fog systems to increase the overall Fog system performance.

\begin{figure}[!t]
    \centering
    \includegraphics[trim=483 195 6 3, clip, width=0.45\textwidth]{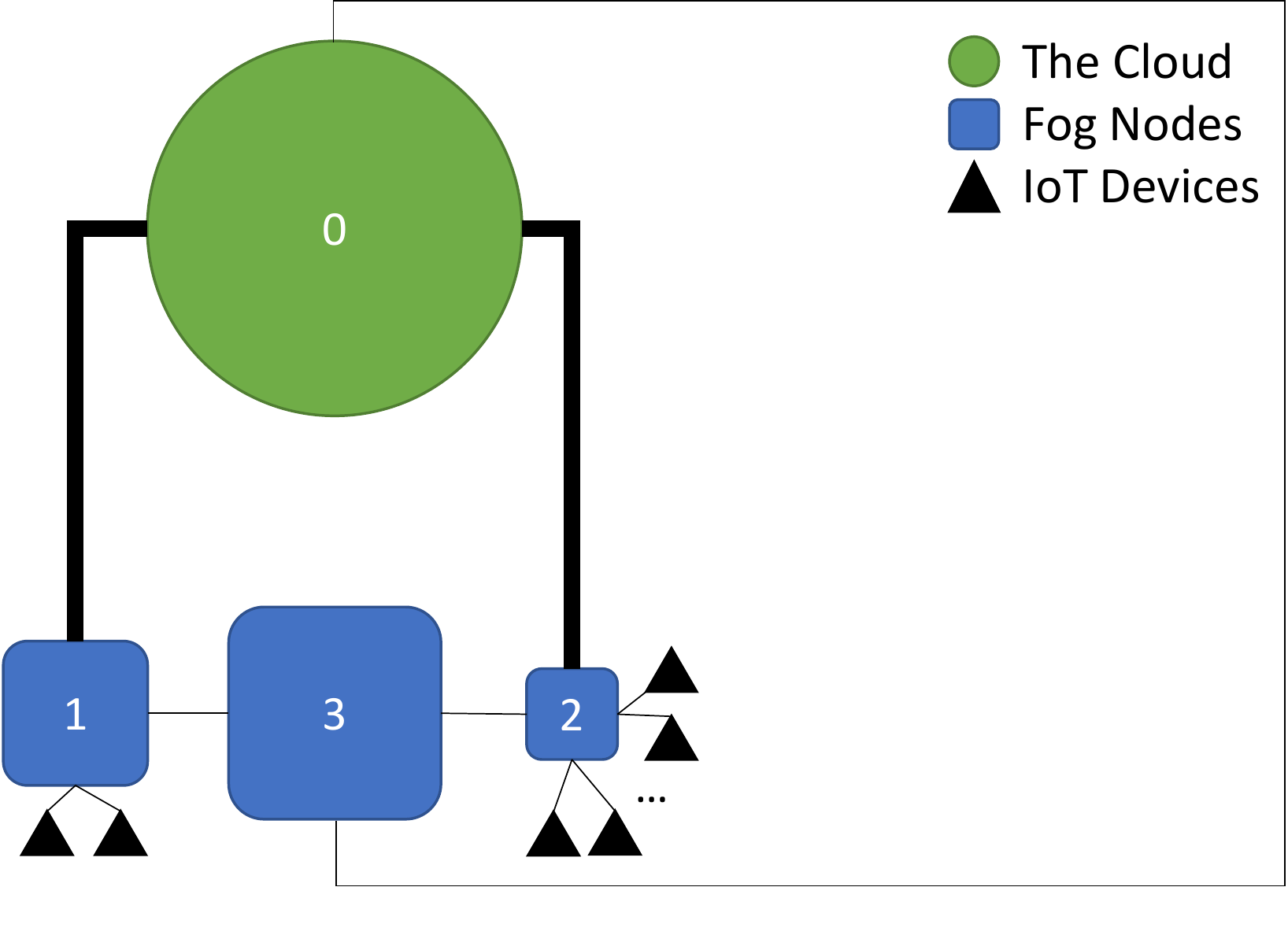}
    \caption{Generic Fog topology that emphasizes the need for load balancing.}
    \label{fig:simpleTopo}
\end{figure}

We also provide in this work a generic network architecture (see Fig. \ref{fig:simpleTopo}) with heterogeneous and unbalanced resources, unbalanced load distribution, heterogeneous workload requirements, and a semi-hierarchical topology. It demonstrates the need for load balancing in unbalanced Fog environments with bottlenecks in computational and communication resources. It is used to evaluate and compare the performance of our approach with other service selection methods. This architecture avoids common simplifications in the literature, like using hierarchical topologies with homogeneous resource and workload requirements \cite{Comparison}.

\begin{figure}[!t]
    \centering
    \includegraphics[trim=240 120 280 120, clip, width=0.45\textwidth]{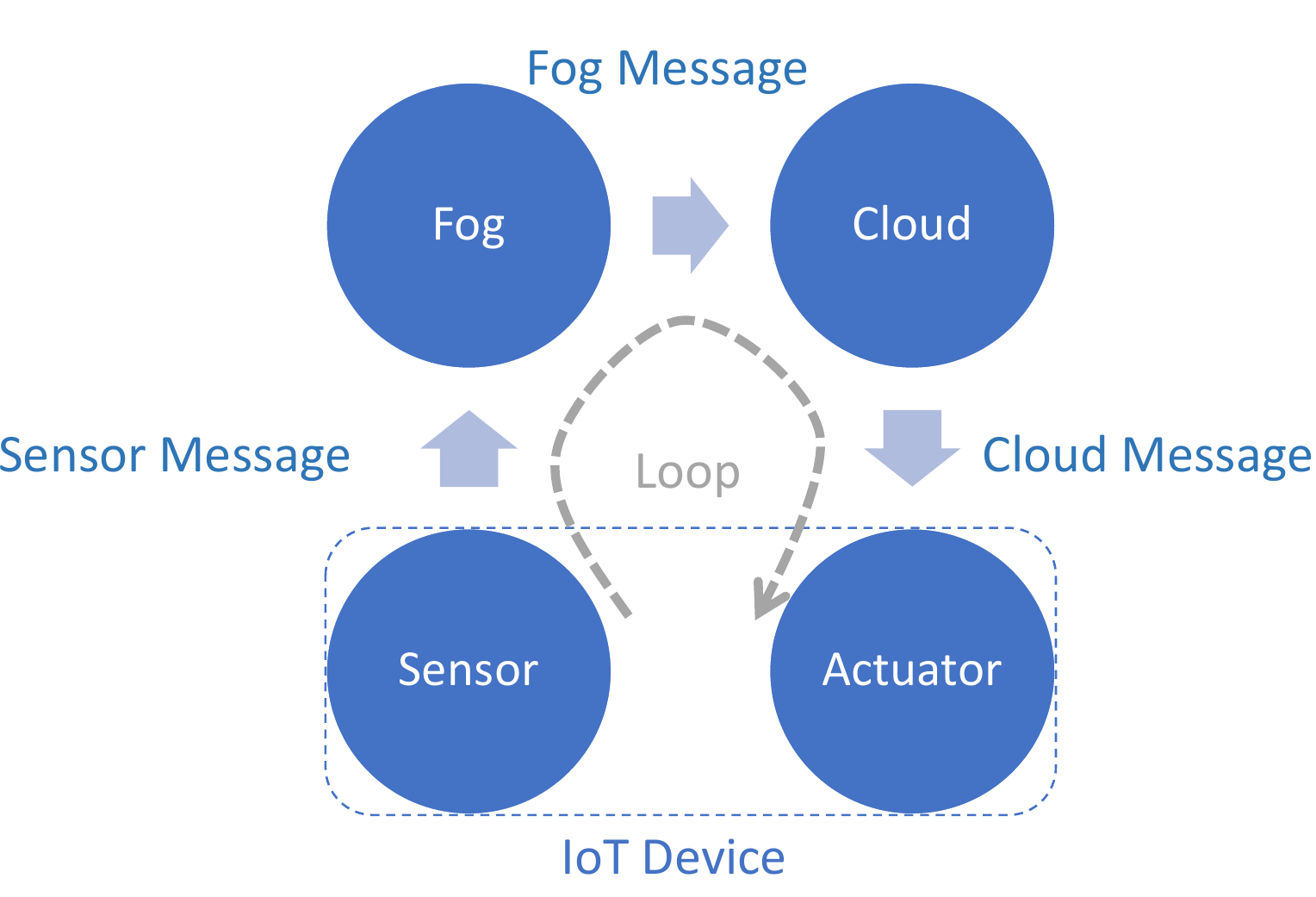}
    \caption{A simple distributed IoT application workflow for the generic Fog architecture, with a single application loop.}
    \label{fig:simpleloop}
\end{figure}

We then evaluate our approach in a more realistic experimentation setup, which consists of multiple Fog nodes that are randomly configured to simulate unbalanced resource and load distribution. This larger setup confirms both the performance of our proposed approach, as well as the effectiveness of our generic architecture in evaluating load balancing algorithms in a simplified setup. In addition, we consider pipelined IoT application workflows with computational modules that span from the edge of the network to the Cloud (see Fig. \ref{fig:simpleloop}). This consideration avoids simplifying Fog problems into Edge problems, where atomic IoT workloads are assigned to a single computing entity that then sends feedback to the source IoT device in response \cite{WorkloadBalancing}.

Hence, our main contributions can be summarized as the:
\begin{enumerate}
    \item Design and implementation of a generic Fog architecture that can be used to evaluate load balancing algorithms away from the common simplifications in the literature.
    \item Design and implementation of ELECTRE MCDA-based load balancing algorithm that simultaneously considers five selection criteria for its decision-making process.
    \item Comparison of our proposed approach with traditional methods that are commonly used in practice, which are wrongly assumed to behave well in Fog environments. 
    \item Evaluation of our proposed load balancing solution in a larger and more realistic setup to confirm our findings.
\end{enumerate}


The rest of the paper is organized as follows. Section \ref{sec:related} presents a comparison between this work and existing work in the literature. We introduce the ELECTRE algorithm in Section \ref{sec:electre}. Section \ref{sec:design} presents the generic architecture used to evaluate the proposed solution. Section \ref{sec:evaluation} discusses the results and the superior performance of our approach. Finally, Section \ref{sec:conclusion} concludes the paper and presents some future directions for the work.

\section{Related Work}
\label{sec:related}
Increasing the performance of IoT applications using Fog resources can be done in two ways: (1) Constructing Fog infrastructures that match the expected IoT traffic in a system \cite{Dimensioning1}. (2) Allocating the resource of existing Fog infrastructures to run IoT applications, and optimally distributing the load of these applications to those resources in real-time. The first approach works for geographical regions that do not have existing Fog infrastructures. The goal here is to find the optimal geographical locations of Fog nodes and to determine their optimal resource requirements, i.e., CPU, memory, and storage. In contrast, the second approach is more realistic as we simply use the resources of existing hardware in the network using optimal resource allocation and load distribution.

Before balancing the load in Fog environments, we need to allocate Fog resources to distributed application modules. This problem is also called the Fog placement problem, i.e., where to deploy service instances of IoT application modules. The allocation of resources can be static or dynamic \cite{AReview}. With static allocation, services do not migrate after their initial deployments, while they can migrate during their lifetime when dynamic allocation is used, which is more realistic in dynamic mobile environments. In this case, efficient mobility-aware bandwidth reservation schemes are required to support real-time IoT applications to preserve the valuable and scarce bandwidth resources \cite{MobilityPrediction2}. In addition, the overhead of migrating services and their associated data can be unacceptable in realistic deployments, which is often underestimated in the literature \cite{MobFogSim}. It can be ignored only when migration is done on micro Fog services that are not associated with user data. 

Velasquez et al. \cite{RankedPlacement}, for example, reduced network latency in Fog environments by optimizing service placement using Integer Linear Programming (ILP). The optimization is based on the popularity of each application, i.e., the number of requests for each application. Then, they proposed a faster, but near-optimal, heuristic solution based on the PageRank algorithm. Their results showed that the heuristic approach was better in balancing the load between Fog nodes with latency values close to those of the ILP optimal solution. They used the YAFS simulator \cite{YAFS} to test their solution and compared their results with the First Fit (FF) placement algorithm. However, they only solved the initial placement problem, i.e., static resource allocation, in Fog environments, which cannot adapt to dynamic environments with mobile nodes. 

The Fog placement problem was also addressed using other optimization algorithms, including Genetic algorithms \cite{Guerrero2018}, Particle Swarm Optimization (PSO) \cite{SDNLB4IoV}, Gravitational Search Algorithm (GSA) \cite{FMEC}, Monte Carlo simulations \cite{7919155}, and linear programming \cite{Velasquez2017}. However, solving the Fog placement problem does not necessarily provide resiliency in the system and it might require service and data migration to support mobile networks, which introduces a huge overhead in dynamic systems. Ni et al. \cite{PetriNets} compared static allocation strategies with dynamic resource allocation in the Fog, where they filtered, classified, and grouped Fog nodes based on their computing capabilities and their credibility. They considered the price and time costs needed to complete the tasks along with the credibility of end-users and Fog resources. However, users or resources can have similar credibility, which can cause sub-optimal behavior when breaking ties. In addition, the dynamic nature of users' and resources' credibility can cause deviations when calculating these values.

Téllez et al. \cite{Tabu} balanced the load between Fog nodes and the Cloud by finding the optimal task allocation using an ILP-based Tabu search method. They transformed multi-objective optimization into a single-objective problem using the Pareto Frontier method. They used a Task Scheduling Coordinator to receive the tasks, schedule them, and assign them to suitable computing nodes. Puthal et al. \cite{SecureLB} used Breadth-First Search to choose the best computing node in the network to improve resource utilization and job response time in the system.

Xu et al. \cite{DynamicRS} balanced the load between Fog nodes based on CPU, RAM, and bandwidth resources using static resource allocation and dynamic service migration. Services were partitioned to Fog nodes based on their type and predefined request generation rates. They compared their approach with FF, Best-Fit (BF), FF-Decreasing (FFD), and BF-Decreasing (BFD). Pereira et al. \cite{ALBALGOFOG} proposed a priority-based load balancing algorithm with two predefined priority levels. They used a centralized controller with a global knowledge of system resources and workload requirements. The controller uses a search table to select the best available computing node and can create more nodes if needed. High priority tasks are sent to the node with the lowest load and latency and the highest number of cores and available memory. While low priority tasks wait for nodes with low load or will be sent to the Cloud.


Pinto Neto et al. \cite{loadDistribution} compared their multi-tenant load distribution algorithm with a delay-driven load distribution strategy. They considered homogeneous Fog environments with redundant services replicas. In their hierarchical architecture, a management node in the upper layer manages all Fog nodes in the lower layer. It keeps a table for the resource utilization in each Fog node as well as the communication delay between adjacent nodes. When a Fog node receives a task, it consults the management node to select the best node to run this task based on its priority and delay requirements. If all nodes are fully loaded, tasks are kept in a queue in the management node. When a node becomes available, the management node sends to it the task with the highest priority.

Despite their near-optimality, machine learning based solutions have been also considered for load balancing problems in Fog systems. Mseddi et al. \cite{ResAllocDQN}, for example, used Deep Q-networks (DQN) to maximize user satisfaction with mobile Fog nodes that are free to join or leave the network. It runs in multiple controllers, where each controller is responsible for a group of Fog nodes in its region. To build the state of the environment, they included the location of Fog nodes and IoT devices and the requirements of computational tasks. Each controller selects the best Fog node in its region that can perform the task within a predefined delay threshold. They compared their results with ILP, random, and nearest node selection algorithms. However, they assumed identical independent tasks with homogeneous workloads, which violates the characteristics of distributed IoT applications.

Talaat et al. \cite{LBOS} also used a Master controller for each isolated Fog region, which continuously monitors traffic, collects resource and load information, and fairly distribute requests to Fog nodes in their regions. They used RL and genetic algorithms to allocate and migrate tasks using Adaptive Weighted-RR (AWRR), where the decisions are based on the CPU, RAM, and Cache values of each node and other predefined conditions. They compared their approach with RR, Weighted-RR (WRR), and Least-Connection (LC) selection methods. Baek et al. \cite{FogRLLB} used SDN controllers to lower the overloading probability of Fog nodes while reducing job latency, which was achieved by offloading the optimal number of tasks to neighboring Fog node. They calculated a load index value using Q-Learning to represent the computational load on a given node based on a global system average. They compared their work with random, nearest, and least-queue node selection schemes.

Wang and Varghese \cite{contextaware} used Cloud-based DQN agents to decide on the number of services that should be offloaded from the Cloud to the Fog layer, and their optimal placement, considering QoS and running costs. They formulated Fog nodes as the environment and nodes' characteristics as the state of the agent. They compared their approach with a static predetermined service distribution. However, they did not consider dynamically changing environments, where the number of users and workload generation rates can change over time. In addition, they did not consider multi-tenancy, where multiple Fog applications are deployed on a single Fog node simultaneously.

Approximation-based approaches provide semi-optimal results through function approximation. Conversely, random-based load balancing approaches provided good performances with simple implementations and minimum resource requirements while avoiding complex coordination between Fog nodes. For instance, Beraldi and Alnuweiri \cite{Randomization} proposed a simple load balancing algorithm that leverages the power of random choice property. In their solution, every time a node receives a job, it sends a request to a randomly selected node to take this task. The selected node informs the task initiator in case it accepts it. Otherwise, it repeats the process sequentially by sending another request to another randomly selected node. This sequential solution is equal to their parallel solution, proposed in \cite{Randomization5}, with a fan-out value equal to 1, which means sending the request to a single random node. 

In another work, Fog nodes probe random nodes to offload tasks if their current load exceeds a predefined threshold \cite{Randomization3}. This is similar to the work in \cite{Randomization4} except for probing a predefined number of random nodes, then selecting the least loaded. Similarly, Beraldi et al. \cite{Randomization2} proposed sequential and adaptive random forwarding algorithms. Here, if the job is already offloaded more than a predefined threshold, it is executed locally or dropped based on the available resources. Otherwise, the node decides between processing the task locally and offloading it to a random node based on its current load. The sequential algorithm uses a predefined load threshold for each node, while the adaptive algorithm tunes this threshold based on the number of times the job was already forwarded.

There are many limitations in existing load balancing approaches in the literature (summarized in Table \ref{tab:compare}). That motivated us to propose a solution that mitigates these limitations. Our proposed load balancing solution optimizes multiple objectives simultaneously using an MCDA-based method to improve the overall system performance. MCDA-based methods were used to make resource selection decisions in different domains, including Mobile Crowd Computing systems, where smart mobile devices are utilized as computing resources \cite{CrowdComputing}. These methods achieved satisfactory performance and provided quality of service by selecting the most suitable resources. Furthermore, MCDA methods were also used to efficiently solve the service selection problem in Cloud Computing environments \cite{CloudSelection}.

\begin{table*}[!t]
\scriptsize
\caption{Comparison between existing load balancing approaches for Fog Networks \label{tab:compare}}
\centering
\begin{tabular}{|p{0.195\textwidth}|p{0.15\textwidth}|p{0.14\textwidth}|p{0.2\textwidth}|p{0.195\textwidth}|}
\hline
\textbf{Approach} & \textbf{Optimization Objectives} & \textbf{Evaluated against} & \textbf{Limitations} & \textbf{How we Address these limitations}\\
\hline
Random (\textit{offloading} \cite{Randomization}\cite{Randomization5}, \textit{probing} \cite{Randomization3}\cite{Randomization4}, \textit{forwarding} \cite{Randomization2}) & - & No load balancing & Can unnecessarily waste resources as it does not guarantee optimality & Provide optimal/near-optimal solution \\
\hline
Popularity Ranked Placement \cite{RankedPlacement} & Minimize network latency & ILP and FF & Initial/static service placement only & Consider dynamic service selection to adjust to system changes \\
\hline
Credibility-based Grouping \cite{PetriNets} & Minimize task completion time and processing price & Predetermined static allocation & Credibility deviation and ties & Address ties by selecting the nearest node to save network resources \\
\hline
ILP-based Tabu search \cite{Tabu} & Minimize computational cost using Fog or Cloud & - & Price-based task assignment only & Consider improving service utilization and execution delay \\
\hline
Breadth-First Search \cite{SecureLB} & Improve resource utilization and job response time & Static, proportional, and random & Uses a balanced Fog architecture & Consider unbalanced Fog topology with unbalanced load distribution \\
\hline
Static allocation with migration \cite{DynamicRS} & Minimize the load-balance variance & FF, BF, FFD, and BFD & Migration cost is not considered & Avoid migration, which depends on network resources and its traffic \\
\hline
Priority-based search \cite{ALBALGOFOG} & Homogeneous tasks distribution & No load balancing & Predefined priorities and thresholds that must be determined by experts & Avoid predefined parameters to dynamically adjust to system changes \\
\hline
Multi-tenant load distribution \cite{loadDistribution} & Priority and communication delay & Delay-driven load distribution & Homogeneous Fog architecture & Consider heterogeneous resources with heterogeneous workloads \\
\hline
DQN allocation and migration \cite{ResAllocDQN} & QoS satisfaction within predefined delay threshold & ILP, random, and nearest & Considers identical atomic tasks & Avoid pre-training/retraining and consider heterogeneous tasks \\
\hline
RL and genetic algorithms \cite{LBOS} & Reduce allocation cost and response time & LC, RR, WRR & Migration cost is not considered & Avoid pre-training/retraining and avoid migration \\
\hline
Q-Learning \cite{FogRLLB} & Lower overloading probability and job latency & Random, nearest, and least-queue & Do not consider unbalanced scenarios & Avoid pre-training/retraining and consider unbalanced scenarios \\
\hline
Dynamic redeployment with DQN \cite{contextaware} & Optimal placement considering QoS \& running cost & Static service distribution & Single Application per Fog node only & Avoid pre-training/retraining and consider multiple applications \\
\hline
\end{tabular}
\end{table*}

The varying significance of the selection criteria makes MCDA methods suitable for load balancing problems in the Fog. ELECTRE, for example, is an effective outranking MCDA method that was used to solve the Cloud service selection \cite{CloudSelection} and the Fog service placement \cite{MCDA} problems. However, to the best of our knowledge, MCDA-based methods have never been used to balance the load in Fog environments using service selection decisions of stateless micro Fog service replicas. To fill this gap, we propose in this work an MCDA-based load balancing solution using the ELECTRE algorithm, where we simultaneously minimize the hop count, propagation delay, processing delay, execution delay, and waiting delay to improve the overall system performance. We compare the performance of our approach against traditional service selection methods, i.e., random, RR, nearest node, and fastest service selection. Evaluating load balancing algorithms against such baselines is a common practice in the literature (see Table \ref{tab:compare}). To advance the state of the art, the proposed solutions must mitigate the limitations of existing approaches in the literature before improving the system performance in terms of execution delay and resource utilization.

Table \ref{tab:compare} shows that our proposed solution addresses a number of limitations in existing work in the literature. For example, our solution provides a near-optimal solution for the load balancing problem in Fog environments. Unlike randomized methods (e.g., \cite{Randomization}\cite{Randomization5}\cite{Randomization3}\cite{Randomization4}\cite{Randomization2}), which do not guarantee optimal load distribution, and can unnecessarily waste compute and network resources. In addition, our proposed approach can easily adapt to dynamic changes in the topology of the system, availability of compute and network resources, as well as the changes in load distribution and its generation rate. In contrast to static allocation solutions (e.g., \cite{RankedPlacement}\cite{DynamicRS}), with initial service placement of Fog services, which can only perform well in unrealistic static environments that never change overtime. Our proposed solution also avoids predefined priorities and thresholds that are common in existing solutions (e.g., \cite{ALBALGOFOG}), which are hard to be estimated even by experts. Avoiding such predetermined parameters allows for dynamic load balancing solutions that can easily adjust to system changes.

We also carefully address the problem of having ties between alternatives, which is a problem in credibility-based grouping approaches (e.g., \cite{PetriNets}). Hence, in case of a tie, i.e., more than one alternative is identified as top-ranked, the nearest Fog node is selected to save network resources. Additionally, our approach aims to improve service utilization in computing nodes while minimizing execution delay; this is different from price-based task assignment solutions (e.g., \cite{Tabu}) that only focus on minimizing the offloading monetary cost without considering about system performance. Moreover, we evaluate our approach in unbalanced Fog architectures with unbalanced load distribution, heterogeneous resources and workloads, and multiple simultaneous applications. In the literature, however, the evaluation is often simplified by considering balanced and homogeneous system architectures (e.g., \cite{SecureLB}\cite{loadDistribution}\cite{FogRLLB}), identical or homogeneous workloads (e.g., \cite{ResAllocDQN}), and even considering a single application service in each Fog node (e.g., \cite{contextaware}).

Considering redundant stateless micro Fog services provide resiliency in the system by maintaining service availability in case of possible failures in Fog nodes or network links. In addition, this also avoids the need for service and data migration during the load balancing and workload offloading process. Avoiding migration also means avoiding its induced cost and overhead on the system, which is often ignored for simplicity in existing contributions in the literature (e.g., \cite{DynamicRS}\cite{LBOS}). Furthermore, the cost of training for RL-based algorithms, and retraining in case of dynamic environments, is also often ignored in the literature (e.g., \cite{ResAllocDQN}\cite{LBOS}\cite{FogRLLB}\cite{contextaware}). This cost is usually huge since RL agents need to train for hundreds of thousands of training steps using powerful GPUs and/or CPUs before reaching an optimal solution. With our approach, no training or retraining is needed as it simply builds a number of matrices to make a service selection decision; this enables our approach to work with a small overhead, especially in dynamically changing environments.

\section{ELECTRE}
\label{sec:electre}

Balancing the load in Fog networks includes minimize network congestion, network latency, and application execution delay while maximizing the resource utilization of all computing nodes in the system. These objectives can be optimized simultaneously using a multi-objective optimization approach, a subcategory of MCDA methods. MCDA methods, also referred to as Multi-Criteria Decision-Making (MCDM) methods, aim to explicitly evaluate multiple conflicting criteria in a decision-making process. Having MCDA methods solve the Cloud service selection problem, the Fog placement problem, and the resource selection problem in Mobile Crowd Computing motivated us to use it to efficiently distribute the load in Fog networks.

ELECTRE, for instance, is a greedy outranking MCDA method that is based on pairwise comparisons between multiple criteria \cite{ELECTRE}. This category of search-based optimization methods can optimize Fog service selection decisions by removing outranked alternatives. It is a search-based method because it works by checking whether one alternative is better or worse than the other using pairwise comparisons between all possible alternatives in the solution space. Search-based methods quickly adapt to dynamically changing environments since they do not require training or predefined parameters. ELECTRE, which is a French acronym for ELimination and Choice Expressing the REality, was first proposed in 1965 by Bernard Roy. Since then, different versions of ELECTRE methods have emerged, including ELECTRE III \cite{ROY}. 

ELECTRE III solves some limitations in ELECTRE II, like dealing with data inaccuracies, imprecision, and uncertainties. In this method, a binary outranking relationship between two alternatives $a$ and $b$ is represented as $aSb$, which means that $a$ is at least as good as $b$. This relationship is not symmetric, and hence four different scenarios exist:
\begin{equation*}
{S(a,b)} = \begin{cases}
aSb \text{ \& } not(bSa) &\Rightarrow a \text{ is preferred to } b\\
not(aSb) \text{ \& } bSa &\Rightarrow b \text{ is preferred to } a\\
aSb \text{ \& } bSa &\Rightarrow \text{indifference}\\
not(aSb) \text{ \& } not(bSa) &\Rightarrow \text{incomparability}
\end{cases}
\end{equation*}

Alternatives are evaluated by $K$ problem-related criteria that are weighted by importance factors $w_i > 0$, where $\sum_i^K{w_i} = 1$. For every criterion $g_i$; $g_i(a) >  g_i(b)$ when alternative $a$ is preferred to $b$ according to that criterion. The novelty of ELECTRE III is the introduction of the concept of pseudo-criteria using discriminating thresholds for every criterion, i.e., indifference $q_i(g_i(a))$ and preference $p_i(g_i(a))$ thresholds. These thresholds can be constant or can dynamically vary along the scale of the values of each criterion, such that $p_i \geq q_i \geq 0$ for every criterion $i$.

ELECTRE III defines a credibility matrix $\sigma(aSb)$ for the outranking relation $aSb$ using a concordance matrix $c(aSb)$ and a discordance matrix $d(aSb)$. This credibility matrix is given as follows: $\sigma(aSb) = c(aSb) \prod_{i=1}^K{T_i(aSb)}$, where:
\begin{equation*}
{T_i(aSb)} = \begin{cases}
\frac{1-d_i(aSb)}{1-c(aSb)}, &{\text{iff }} d_i(aSb) > c(aSb)\\
1, &{\text{otherwise.}}
\end{cases}
\end{equation*}

$c(aSb)$ is defined as $\sum_i^K{w_i c_i(aSb)}$, where:
\begin{equation*}
{c_i(aSb)} = \begin{cases}
1 &{\text{if }} g_i(b) \leq g_i(a) + q_i(g_i(a))\\
0 &{\text{if }} g_i(b) < g_i(a) - p_i(g_i(a))\\
\frac{g_i(a) - g_i(b) + p_i(g_i(a))}{p_i(g_i(a)) - q_i(g_i(a)) } &{\text{otherwise,}}
\end{cases}
\end{equation*}

\noindent while the discordance matrix is defined as:
\begin{equation*}
{d_i(aSb)} = \begin{cases}
1 &{\text{if }} g_i(b) > g_i(a) + v_i(g_i(a))\\
0 &{\text{if }} g_i(b) \leq g_i(a) + p_i(g_i(a))\\
\frac{g_i(b) - g_i(a) - p_i(g_i(a))}{v_i(g_i(a)) - p_i(g_i(a))} &{\text{otherwise.}}
\end{cases}
\end{equation*}

The concordance represents the majority among all different criteria that favor an alternative to another one. While the discordance represents, when the concordance condition holds, the minority criteria that strongly oppose that assertion. 

Also, ELECTRE III uses the concept of veto thresholds for each criteria, i.e., $v_i(g_i(a))$, which is a variable threshold such that $v_i > p_i$ for every criterion $i$. This threshold represents the power of a given criterion to be against that assertion when the performance difference between the two alternatives for that criterion is greater than this threshold. In ELECTRE III, veto thresholds are used to define the discordance matrix, hence, they can be set large enough to make the discordance values equal to zero, and hence, making $\sigma(aSb) = c(aSb)$.

At least three criteria are needed to build an ELECTRE-based decision model. However, ELECTRE methods are more adequate when there are between five and thirteen criteria. Also, the criteria should be heterogeneous with ordinal, or weakly interval, scales, where the loss of one criterion does not benefit another. In addition, small differences are eliminated with the help of indifference and preference thresholds that dynamically scale to the data.

\section{System Design}
\label{sec:design}

To formulate our Fog environment, we use capital letters to refer to the set of items, i.e., nodes, links, modules, or messages. To refer to a single item in that set we use the lower case of that letter with a subscript. Fog environments are composed of a set of $N$ nodes, i.e., computing nodes ($N_C$ Cloud nodes and $N_F$ Fog nodes) and non-computing nodes ($N_{IoT}$ IoT nodes and $N_O$ other nodes), where node $n_x$ is defined by its compute ($IPT_x$) and memory ($RAM_x$) resources, as shown below:
\begin{align*}\label{eq:N}
    N &= N_C \cup N_F \cup N_{IoT} \cup N_O \\
    &= n_1, n_2, n_3, \cdots, n_z, \quad \text{where} \enspace n_x \defeq \langle IPT_x, RAM_x \rangle
\end{align*}

$N_{IoT}$ represents sensors and actuators with pure source and sink nodes, respectively. $N_O$ represents routers, switches, proxy servers, and firewalls. For load balancing algorithms, nodes must be characterized by identifiers (IDs), the number of instructions performed per unit of time (IPT), and their memory capacity (RAM). These nodes are connected through $L$ wired/wireless bidirectional links, where each link $l_x$ is characterized by the pair ($n_i$, $n_j$) that it connects, its bandwidth $BW_x$, and its propagation delay $PR_x$:
\begin{equation*}\label{eq:L}
    L = l_1, l_2, l_3, \cdots, l_z , \quad \text{where} \enspace l_x \defeq \langle n_i, n_j, BW_x, PR_x \rangle
\end{equation*}

IoT distributed applications are represented using distributed data flow (DDF) models \cite{7356560}. There can be several distributed applications simultaneously running in the system; let $A$ be the set of these applications. The workflow of each application $a_x \in A$ is represented by a set of modules $M_x$ and a set of dependencies between these modules $D_x$, as shown here:
\begin{equation*}\label{eq:App}
    A = a_1, a_2, a_3, \cdots, a_z , \quad \text{where} \enspace a_x \defeq \langle M_x, D_x \rangle
\end{equation*}

Application modules can be categorized into three main classes, i.e., computing modules ($M_C$), pure source modules ($M_{SRC}$), and pure sink modules ($M_{SNK}$):
\begin{align*}\label{eq:M}
    M &= M_C \cup M_{SRC} \cup M_{SNK} \\
    &= m_1, m_2, m_3, \cdots, m_z, \quad \text{where} \enspace m_x \defeq \langle RAM_x \rangle
\end{align*}

Pure source/sink modules are implemented in IoT nodes as they do not perform computations on the generated/consumed data. Computing modules, in contrast, are deployed in computing nodes, i.e., Cloud, Fog, and Edge nodes. They serve incoming workloads and often generate a new workload for each processed request. They can also act as non-pure source and sink modules by periodically emitting aggregate or sync information based on the data they receive over a certain period. In addition, they can collect data without emitting messages in response, i.e., storing data for future analysis. 

Application workflows are represented as a Directed Acyclic Graph (DAG), where nodes represent the modules and edges represent the dependencies between those modules. Message transfer between module $m_i$ and module $m_j$ represents the dependency, denoted by $d_x$, between these modules. Each message, i.e., IoT workload, is characterized by the number of instructions ($I_x$) required by the workload from the compute entity and the size of the message in bytes ($B_x$):
\begin{equation*}\label{eq:D}
    D = d_1, d_2, d_3, \cdots, d_z , \quad \text{where} \enspace d_x \defeq \langle m_i, m_j, I_x, B_x \rangle
\end{equation*}

Now that we formally formulated our nodes, links, and applications, we can define our environment ($E$) by the set of nodes $N$, the set of links $L$, and the set of applications $A$, as follows:
\begin{equation*}\label{eq:E}
    E \defeq \langle N, L, A \rangle
\end{equation*}

For a realistic evaluation of our proposed approach, we provide a generic Fog architecture (see Fig. \ref{fig:simpleTopo}) to demonstrate the need for load balancing through network bottlenecks, unbalanced Fog resources, and unbalanced workload distribution. Then, we use a more realistic Fog topology to evaluate our proposed approach, with more Fog nodes that are randomly assigned resources under the constrain of creating unbalanced architectures. In both those evaluation scenarios, we consider heterogeneous computation and communication resources with heterogeneous workload requirements to mimic realistic Fog environments. We also consider non-hierarchical architectures to mimic flat Fog systems instead of layered ones. 

Starting with the simple generic architecture in Fig. \ref{fig:simpleTopo}, $Fog_2$ acts as a computation bottleneck as it has the least amount of computational resources, represented by the smallest square in the figure, while being directly connected to the largest number of IoT devices. In addition, the longest communication link, i.e., higher propagation delay, is the one that connects the fastest Fog node ($Fog_3$) with the Cloud. This link has the lower bandwidth, i.e., represented by a thinner line, compared to the links that connect the Cloud with the other two Fog nodes. Hence, this link is a communication bottleneck when $Fog_3$ is used to process IoT workloads.

\begin{figure*}[!t]
\centering
\includegraphics[trim=33 33 0 33, clip, width=\textwidth]{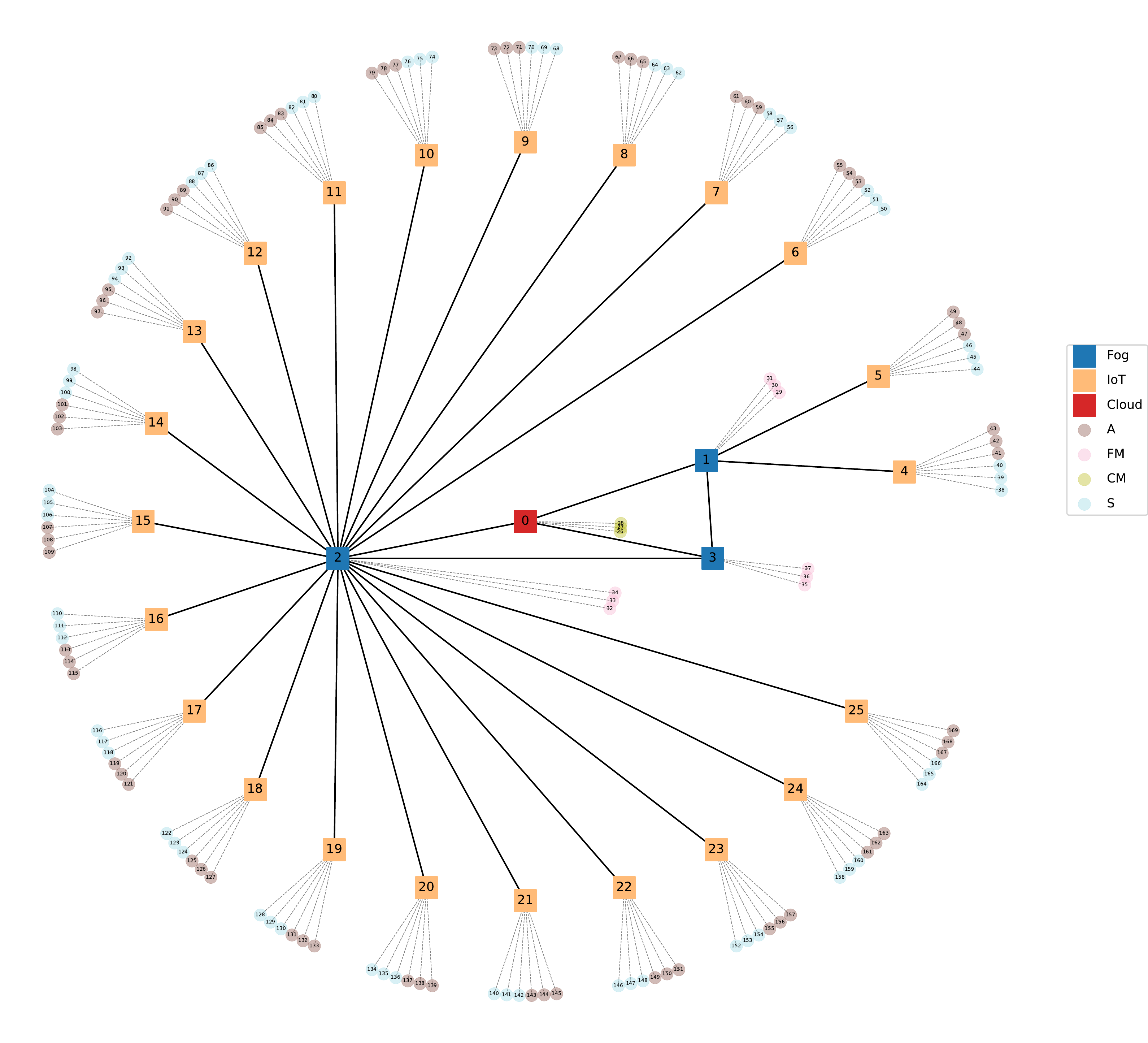}
\caption{A simple generic Fog architecture with three interconnected Fog nodes.}
\label{fig:topo}
\end{figure*}

Fig. \ref{fig:topo} shows a simple implementation of this generic architecture using YAFS \cite{YAFS}, a Discrete-event Simulator (DES) that mimics realistic deployment of Fog-based IoT applications. In this implementation, the Cloud is connected to three interconnected Fog nodes with heterogeneous resources. We simulate the Cloud with one order of magnitude more resources than $Fog_3$, which has one order of magnitude more resources than $Fog_1$, that also has an order of magnitude more resources than $Fog_2$.

A communication bottleneck exists between the fastest Fog node, i.e., $Fog_3$, and the Cloud. In addition, the slowest Fog node, i.e., $Fog_2$, is connected to an order of magnitude more IoT devices than $Fog_1$. While the fastest Fog node only accepts offloaded workloads from the other two Fog nodes since it is not connected to any IoT devices. This unbalanced resource distribution creates a computation bottleneck, which, alongside the communication bottleneck, emphasizes the need for load balancing in this architecture.

In this generic architecture, three applications with heterogeneous workload requirements are deployed in the system, following the distributed application workflow shown in Fig. \ref{fig:simpleloop}. It has four modules with three dependencies, i.e., messages, requests, or workloads, between them. Service replicas for Sensor and Actuator modules are deployed in IoT devices. A single service replica for the Cloud module is deployed in the Cloud, and each Fog node has a stateless service replica of the Fog module. This is a local view of the network, for the load balancer, with the available Fog nodes that can serve a given workload.

Each IoT device has instances of the three Sensor and three Actuator modules, one for each application. Fog nodes and the Cloud have instances of the three Fog and Cloud modules, respectively, one for each application. Workload requirements were chosen relatively to the resources of computing nodes to simulate resource-demanding, moderate, and light workloads. This was done by defining the workload requirements of $App_3$ with one order of magnitude more compute instructions than $App_2$ to simulate resource-demanding and moderate workloads, respectively. In addition, $App_1$ was defined with one order of magnitude less workload requirements than $App_2$ to represent light workloads.

Sensor messages are generated by IoT devices using a Poisson Point Process. These messages wait for their turn to access network links since nodes and links are modeled in YAFS as M/M/1 queuing models. Modeling resources with infinite queue sizes allows studying network saturation as a performance measure instead of workload dropping probability. The next messages in the workflow, i.e., Fog messages, are triggered after Fog nodes are done serving Sensor messages. The same happens when the Cloud finishes processing Fog messages, when it triggers Cloud messages as a response. Cloud messages are finally consumed by the Actuator module in the IoT device that initiated that particular loop.

Load balancing takes place by selecting the proper Fog node to serve an incoming IoT workload. Selecting the nearest Fog node to the source IoT device creates a computation bottleneck on $Fog_2$, which has the least amount of resources while being connected to around 91\% of the system's IoT devices. In contrast, selecting the fastest node creates communication bottlenecks on the links connecting $Fog_3$, because they have smaller bandwidths and larger propagation delays. 

Another simple approach is for each IoT device to select between the three Fog nodes in a Round-Robin manner, which we called Distributed Round-Robin (DRR). We also included a random service selection approach, which leverages the power of random choice property to randomly select a Fog node for every newly generated workload. This approach is similar to the sequential randomization load balancing solution proposed in \cite{Randomization} with the difference that the selected node must accept the assigned request. Hence, the process of informing the task initiator about accepting or rejecting the request, and the process of repeating the selection process in case of a rejection are no longer needed. This helps us study workload saturation rather than studying the dropping probability of workloads.

Our proposed ELECTRE-based method, however, considers five criteria (see Table \ref{tab:criteria}) to select an optimal Fog node, i.e., alternative, to process each generated workload based on the outranking relation $aSb$. Each criteria $i$ for each two alternatives, i.e., $g_i(a)$ \& $g_i(b)$, is represented by the concordance and discordance matrices $c_i(aSb)$ and $d_i(aSb)$, respectively. The objective is to find alternatives that minimize these criteria to determine their outranking relationships. Alternatives that are dominated by others are then identified and eliminated based on these relationships; this results in a smaller set of alternatives. In an iterative procedure, this results in an ordering/ranking of alternatives with the possibility of having ties in the ranks.

The shortest path between the source and destination is used for calculating the hop count, propagation time, and execution delay. If there are multiple shortest paths, i.e., same number of links, the path with the smallest propagation time is selected. Given a request $r$ to travel $\mathbb{L}$ links between its source $s$ and a destination candidate Fog node $x$, along with $\mathbb{T}$ tasks that are currently waiting to be processed in this candidate Fog node, we can formally define these five criteria as follows:

\begin{align*}
    C_{s,x}^{H} & = \sum_{l=0}^\mathbb{L} 1 \\
    D_{s,x}^{PR} & = \sum_{l=0}^\mathbb{L} D_l^{PR} \\
    D_{x,r}^{P} & = \frac{I_r}{IPT_x} \\
    D_{s,x,r}^{E} & = D_{x,r}^{P} + D_{s,x}^{PR} + \sum_{l=0}^\mathbb{L} D_{l,r}^T , \quad \text{where} \enspace D_{l,r}^T = \frac{S_r}{BW_l} \\
    D_x^W & = \sum_{r=0}^\mathbb{T} D_{x,r}^P
\end{align*}

\begin{table*}[!t]
\caption{The criteria used in our ELECTRE-based load balancing algorithm for each candidate Fog node \label{tab:criteria}}
\centering
\begin{tabular}{|c||c|}
\hline
\textbf{Criteria} & \textbf{Definition}\\
\hline
Hop Count $C^{H}$ & The number of links on the path between the node that generates the load and the candidate Fog node. \\
\hline
Propagation Delay $D^{PR}$ & The accumulative propagation time of all the links that connect the workload source and the candidate Fog node. \\
\hline
Processing Delay $D^{P}$ & The time a candidate Fog node takes to process a single computational instruction. \\
\hline
Execution Delay $D^{E}$ & The total execution delay to process the given workload on the the candidate Fog node. \\
\hline
Waiting Delay $D^W$ & The current load on the candidate Fog node. \\
\hline
\end{tabular}
\end{table*}

The link propagation delay is a constant value for each link representing the time required to transmit a single bit from one end of the link to the other. It represents the physical link proprieties and the distance between the pair of nodes it connects, i.e., material and length, respectively. The link communication delay is determined by the transmission delay ($D_{l,r}^T$) of the request $r$, which is given by dividing the message size in bytes ($S_r$) by the link bandwidth in Bps ($BW_l$). This represents the time required to push the whole message into the communication link. Both propagation and transmission delays represent the time required for the whole message to reach the other side of the link. 

The execution delay is calculated based on the network communication delay and the processing time ($D_{x,r}^P$) of request $r$ on the candidate Fog node $x$. The network communication delay is given by adding propagation and transmission delays for every link between the source and the destination. The service time, i.e., processing delay ($D^{P}$), is calculated by dividing the workload's required instructions ($I_r$) by the computational capabilities of the candidate Fog node, i.e., $IPT_x$. Finally, the expected waiting delay ($D_x^W$) in a computing node ($x$) represents the load information for that node. It is calculated by adding the processing delay ($D_x^P$) of every task that is currently waiting to be served by node ($x$).

Except for the expected waiting delay, all other criteria need information about the resource capabilities of candidate Fog nodes, which can be collected using triggered updates only, i.e., during upgrades or downgrades of these resources. However, the current load in Fog nodes, i.e., their resource availability, requires an active monitoring system that collects this information from candidate Fog nodes. This real-time information allows adapting to recurring topological changes and fluctuations in workload generation rates.

The criteria, in our simulations, have identical weights $w_i=1/K=1/5=0.2$ for every criterion $i$, where $K$ is the number of criteria used in the algorithm. This gives equal importance for every criterion since, in ELECTRE, the weights are a measure of relative importance of the criteria \cite{Rogers2000, ROGERS1998552}. These weights do not depend neither on the ranges nor the scales of the values in each criteria \cite{greco2016multiple}. In this case, weights can be viewed as the number of votes given to a criterion in a voting procedure, which indicates the relative importance of each criterion \cite{vincke1992multicriteria, Figueira2010}. The algorithm avoids alternatives, i.e., Fog nodes, that maximize these criteria, which means that; indeed, it chooses alternatives that minimize all criteria for every incoming workload. Therefore, ELECTRE can simultaneously optimize these criteria, given their equal importance, to optimize the overall system performance.

The algorithm dynamically calculates the preference and indifference thresholds, in every decision epoch, using percentile ranking. The veto threshold, for example, was set to the 100\textsuperscript{th} percentile rank of every criterion to avoid discarding alternatives that surpass that threshold. On the other hand, we use an indifference threshold of one-third of the 10\textsuperscript{th} percentile rank while having the preference threshold equal to the 20\textsuperscript{th} percentile rank. This was done to avoid considering small differences of preferences as significant.

To rank Fog nodes from best to worst, ELECTRE outranks the worst alternatives until the best alternative is identified, according to the thresholds discussed above. However, because of indifference thresholds, the possibility of having ties among alternatives exists, and hence a set of alternatives can be identified as top-ranked instead of a single one. In case of a tie, we choose the nearest Fog node to the source device to avoid unnecessary utilization of network links. In 30 experiment runs, the average number of ties in the generic architecture was $15.35 \pm 0.1 \%$, and the nearest Fog node to the workload source was selected instead. 

\begin{figure}[!t]
\centering
\includegraphics[trim=0 200 0 200, clip, width=0.48\textwidth]{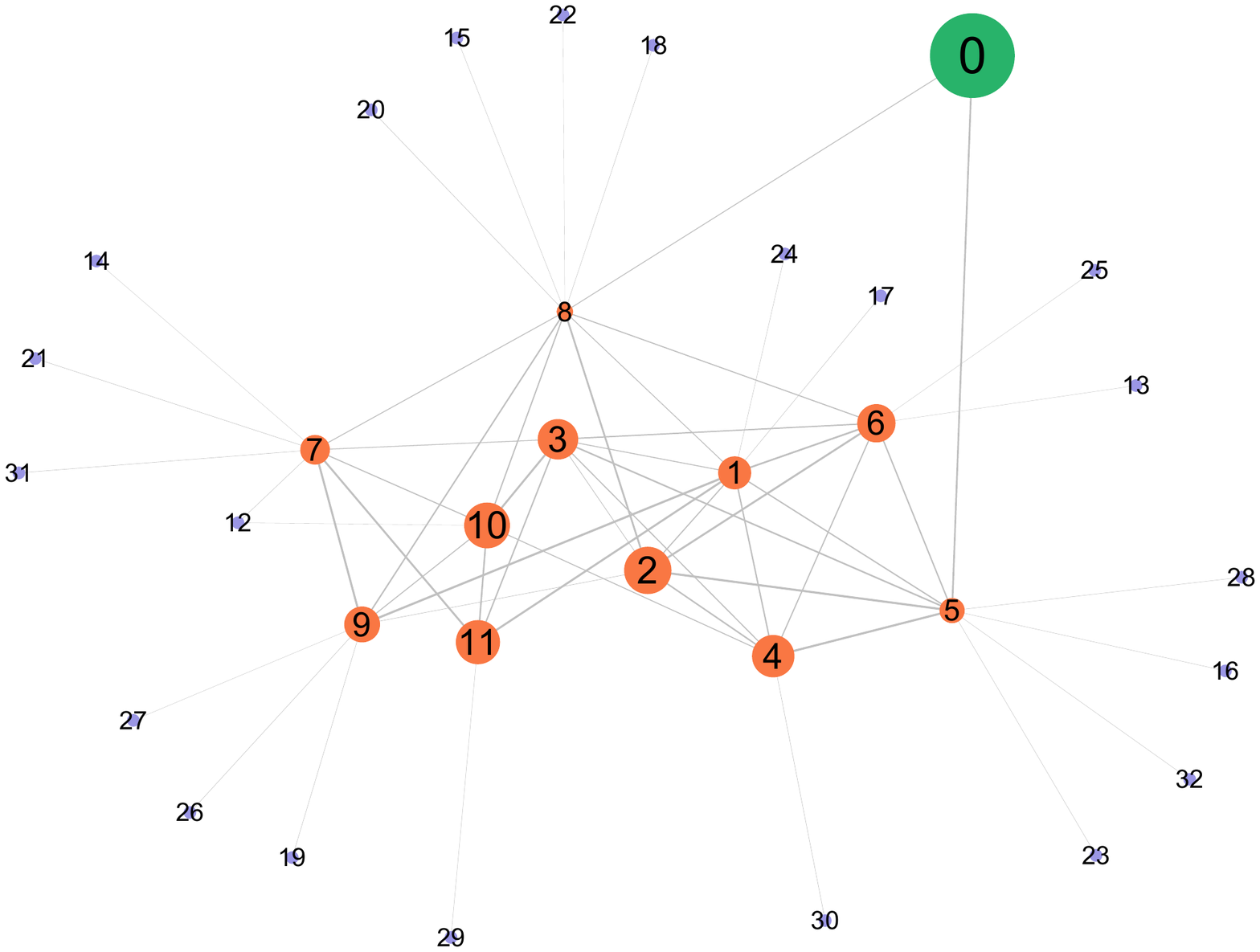}
\caption{Internet Autonomous System (AS) inspired randomized unbalanced Fog environment with 21 IoT devices (purple), 11 Fog nodes (orange), and a Cloud (green). The size of the node represents its compute capabilities.}
\label{fig:topo2}
\end{figure}

To validate our findings from the simple generic Fog architecture, a larger and more complex Fog architecture is used. Fig. \ref{fig:topo2} shows a YAFS implementation of such architecture, which is produced from a randomized graph generator that simulates the Internet Autonomous System (AS) network \cite{networkx}. To create a Fog system out of this graph, we compute the shortest-path betweenness centrality for each node; it is a measure of centrality in the graph based on shortest paths. In other words, for each node, the betweenness centrality is the number of the shortest paths that pass through this node \cite{betweenness}.

To add a Cloud in the center of the architecture, we create a Cloud node and connect it to the two nodes that have the highest betweenness centrality in this randomized graph. Nodes with zero centrality, i.e., edge nodes in the graph, are identified as IoT devices, while the rest of the nodes are identified as Fog nodes. Fog nodes are then assigned their compute resources based on their betweenness centrality to resemble unbalanced resources with unbalanced load distribution. To do this, Fog nodes with high centrality get smaller IPT values compared to Fog nodes with smaller centrality.

\begin{figure}[!t]
    \centering
    \includegraphics[trim=340 130 380 140, clip, width=0.3\textwidth]{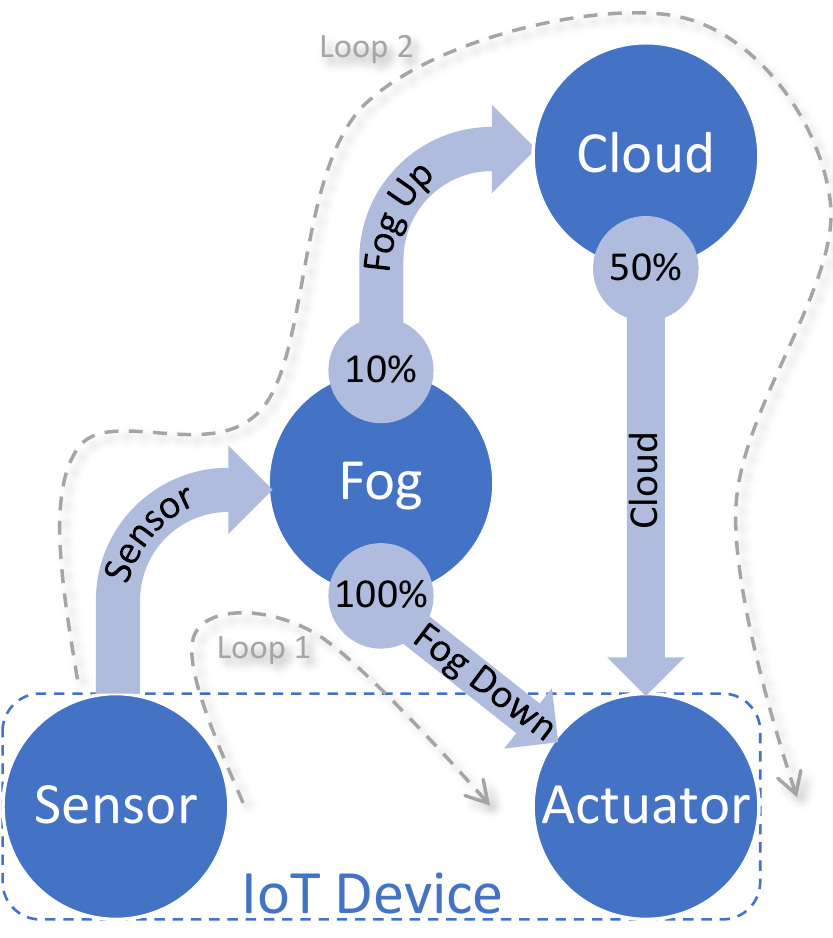}
    \caption{A complex distributed application workflow for the AS-inspired Fog architecture, with two application loops.}
    \label{fig:2loops}
\end{figure}

In addition to the use of a more complex Fog architecture to validate the performance of our approach, we also use a more complex application workflow to execute in these complex Fog environments. Instead of using a single application loop for each application workflow, we now consider two loops for each application workflow (see Fig. \ref{fig:2loops}). Loop 1 represents the immediate feedback, through what we called Fog Down messages. These messages travel from Fog nodes to IoT devices after processing Sensor workloads, where 100\% of Sensor workloads trigger this feedback. 

On the other hand, Loop 2 represents data aggregation and Cloud feedback through Cloud messages, where only 10\% of Sensor workloads will trigger messages to the Cloud (Fog Up messages) and only 50\% of those Fog Up messages will trigger Cloud feedback to the initiating IoT device. This way we mimic the workflow of many realistic applications, where immediate feedback is needed from Fog nodes for every processed Sensor workload. While the Cloud is only involved on a portion of those processed messages to perform data aggregation and/or providing feedback based on the aggregated data. This distributed application workflow reasonably represents existing IoT applications, such as online games, IoV applications, and health monitoring systems.

\begin{figure}[!t]
    \centering
    \includegraphics[width=0.4\textwidth]{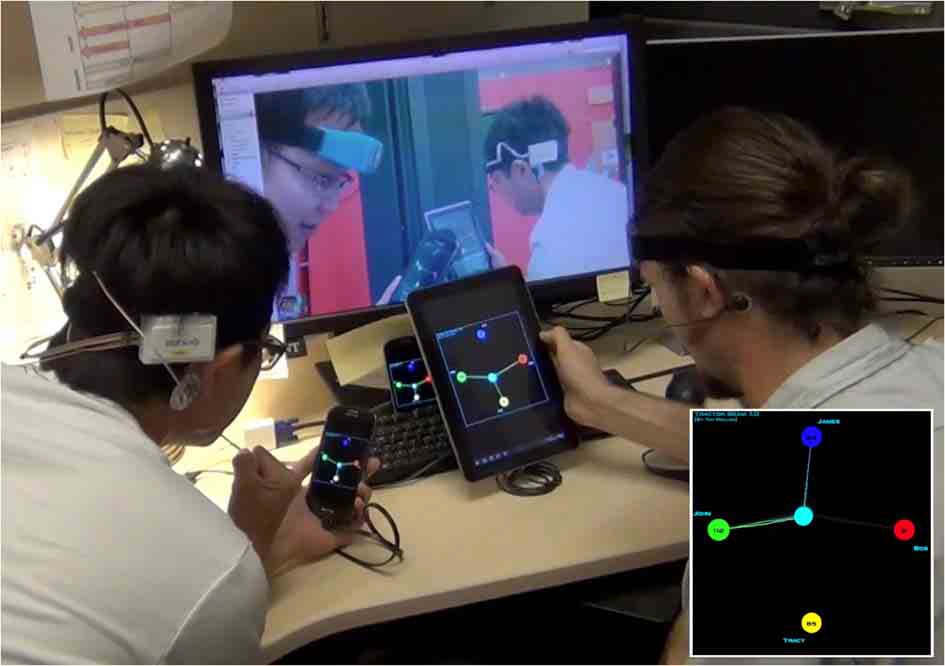}
    \caption{EEG Tractor Beam game session with four people playing over the Internet \cite{EEG}.}
    \label{fig:EEG}
\end{figure}

\begin{figure}[!t]
    \centering
    \includegraphics[trim=0 0 340 0, clip, width=0.4\textwidth]{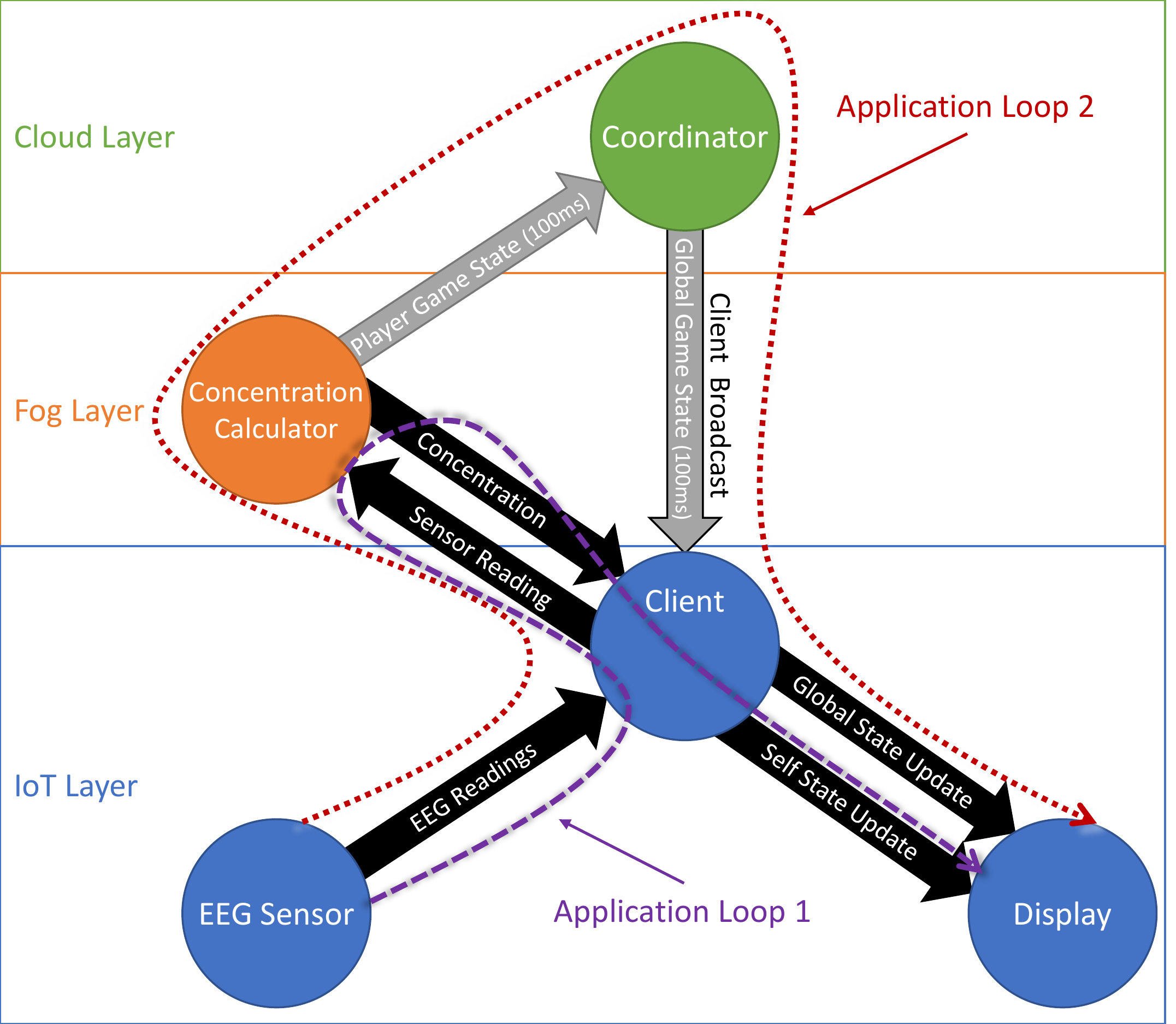}
    \caption{Distributed application workflow for the EEG VR game.}
    \label{fig:eeg_wf}
\end{figure}

The AS-inspired architecture with this distributed application workflow mimics realistic online Virtual-Reality (VR) games, for example, Electroencephalogram (EEG) online game \cite{iFogSim}. In this game, groups of players play collaboratively over the Internet using their brain signals (see Fig. \ref{fig:EEG}). This game is composed of five distributed application modules with seven dependencies (see Fig. \ref{fig:eeg_wf}). players have EEG sensors that send brain signals to their devices, which are pre-processed in their local client modules before being sent to the concentration-calculator module in the Fog node. Fog modules in Fog nodes, i.e. resource-rich gateways, send concentration levels back to the clients to display the status of the player using their display modules. In addition, Fog modules periodically send the state of each player to the coordinator module in the Cloud, which periodically broadcasts the global game state to all client modules.

\section{Evaluation of ELECTRE Method}
\label{sec:evaluation}

\begin{figure}[!t]
    \centering
    \includegraphics[width=0.48\textwidth]{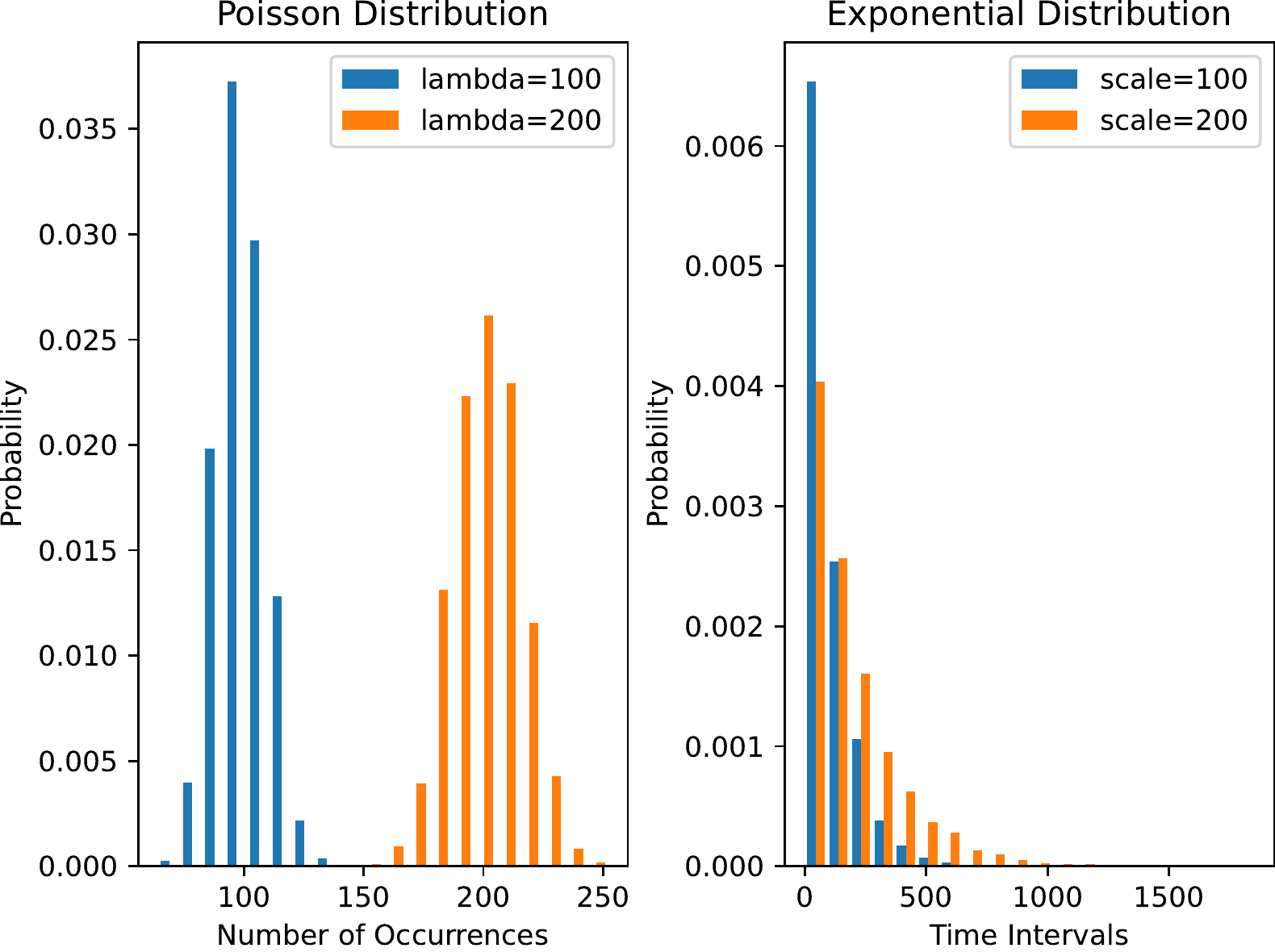}
    \caption{Poisson vs. Exponential Distributions to represent a Poisson-point-process over $10^4$ time-steps.}
    \label{fig:poisson}
\end{figure}

In Table \ref{tab:params}, we show the system's parameters used to perform our experiments on the generic Fog architecture. In these experiments, IoT workloads, i.e., Sensor messages, are generated as a Poisson Point Process using an exponential distribution with a scale parameter $\beta$ of 100, which is common for IoT workload generation rates \cite{MCDA}. The scale parameter is the inverse of the rate parameter $\lambda$, which is another widely used parameter for the exponential distribution that represents the rate of occurrences in a Poisson process. The exponential distribution is the probability distribution of the time between events in a Poisson Point Process (see Fig. \ref{fig:poisson}), i.e., a process in which events occur continuously and independently at a constant average rate \cite{probability}.

\begin{table*}[!t]
\caption{Simulation system parameters for the generic architecture. \label{tab:params}}
\centering
\begin{tabular}{|l|l||l|}
\hline
\multicolumn{2}{|c||}{\textbf{Parameters}} & \textbf{Details}\\
\hline
Algorithms & Service Selection & Random, DRR, nearest, fastest, \& ELECTRE \\
\hline
\multirow{2}{5em}{Simulation} & Number of Iterations & 10 experiment runs each with a different random seed \\
& Simulation Duration & 10000, 100000, \& 1000000 Time-steps \\
\hline
\multirow{3}{5.5em}{Applications} & Number of Apps & 3 Heterogeneous Apps with three modules in each App, i.e., Source, Fog, \& Cloud modules \\
& Workload Instructions & 100, 1000, \& 10000 IPT for each App \\
& Workload Sizes & 10, 100, \& 1000 Bytes for each App \\
\hline
\multirow{3}{4em}{Nodes} & IoT & 22 devices with CPU power of 10 IPT each \\
& Fog & 3 nodes with CPU power of 1000, 10000, \& 100000 IPT for $Fog_2$, $Fog_1$, \& $Fog_3$, respectively \\
& Cloud & 1 Cloud with CPU power of 1000000 IPT \\
\hline
\multirow{3}{4em}{Links} & IoT-Fog & $BW$=1000 \& $PR$=1\\
& $Fog_1$-$Fog_3$ \& $Fog_2$-$Fog_3$ & $BW$=1000 \& $PR$=2 \\
& Fog-Cloud & $Fog_1$-Cloud \& $Fog_2$-Cloud ($BW$=100000 \& $PR$=10), \& $Fog_3$-Cloud ($BW$=1000 \& $PR$=20)  \\ 
\hline
\end{tabular}
\end{table*}

With exponential distribution, the average time between events, i.e., average inter-arrival time, is constant and known through its scale parameter (100 time-steps in our simulations). This distribution is supported on the interval $[0, \infty)$, but 0 is replaced by 1 in our simulations to avoid generating simultaneous tasks in the DES environment, i.e., to create gap intervals between events. Exponential distribution is selected because it easily simulates, using a DES environment, the expected time interval between generated workloads in a Poisson process. Unlike Poisson distribution, which simulates the expected number of workloads that are generated between two time-steps in that process (see Fig. \ref{fig:poisson}). Figure \ref{fig:poisson} shows that smaller intervals are generated with a higher probability; this means that most workloads are generated with small inter-arrival intervals. This is why exponential distribution realistically mimics real-time IoT applications with heterogeneous, but frequent, generation rates.

In our experiments, simulation time-steps refer to seconds for easier analogy and they will be used interchangeably throughout the rest of the text. For instance, the processing power in Table \ref{tab:params} is measured in Instructions Per Time-step (IPT) or Instructions per Second (IPS). Similarly, the bandwidth ($BW$) and the propagation delay ($PR$) are defined in our simulations in Bytes per second (Bps) and seconds, respectively. The values for the bandwidth and the propagation delay in our simulations were chosen to simulate available link resources in shared, i.e., public, network environments, especially at peak hours. 

In the AS-inspired evaluation environment, workloads are generated using the same exponential distribution. However, a different network topology is generated, in each of the 10 experiment runs, using a random AS graph generator. PR and BW resources, for every link in the network, are uniformly chosen from the range of values listed in Table \ref{tab:params2}. These network parameters are inspired from\cite{MCDA}, and they were chosen relatively to workload requirements to simulate resource-demanding workloads in networks with scarce resources. Considering scarce resources helps demonstrate the effectiveness of load balancing algorithms in extreme conditions. Likewise, the resource capabilities of computing nodes and the resource requirements of application workloads were chosen relatively to each other to simulate high-demand workloads that require large computing power from limited computing resources.

\begin{table}[!t]
\caption{Network parameters for the randomly generated AS-inspired architectures \label{tab:params2}}
\centering
\begin{tabular}{|l||p{0.33\textwidth}|}
\hline
\textbf{Parameter} & \textbf{Values}\\
\hline
IoT devices & Nodes with 0 betweenness centrality, each with CPU power of 10 IPT \\
\hline
Fog nodes & IPT values are evenly spaced integers over the interval $[10^3, 10^5]$, values are then assigned to Fog nodes inversely proportional to their betweenness centrality \\
\hline
The Cloud & Connected to the two Fog nodes with the highest betweenness centrality, and assigned CPU power of $10^6$ IPT \\
\hline
\hline
IoT-Fog links & BW and PR values are drawn from a uniform distribution over the intervals $[10^2, 10^3)$ and $[1, 2)$, respectively \\
\hline
Fog-Fog links & BW and PR values are drawn from a uniform distribution over the intervals $[10^3, 10^4)$ and $[2, 4)$, respectively  \\
\hline
Fog-Cloud links & BW and PR values are drawn from a uniform distribution over the intervals $[10^3, 10^4)$ and $[10, 20)$, respectively  \\
\hline
\end{tabular}
\end{table}

\begin{figure*}[!t]
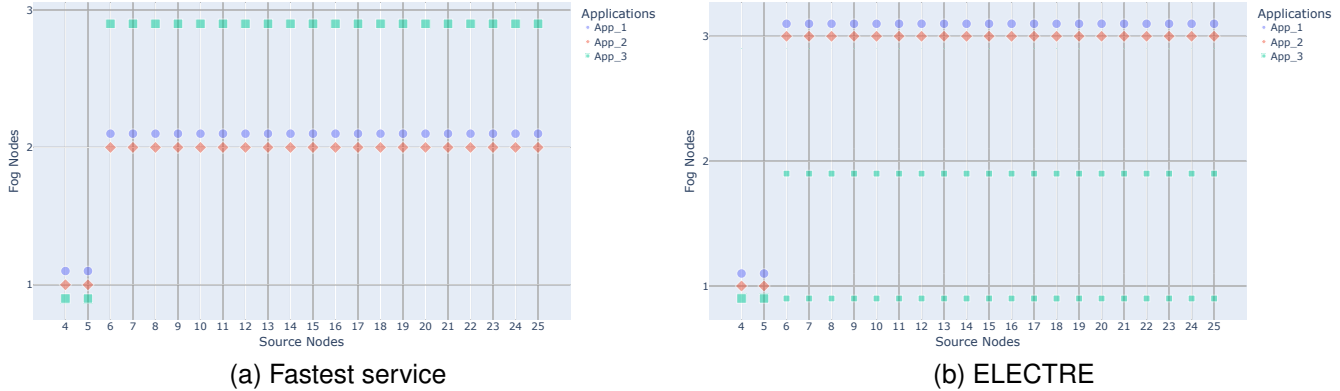

\centering
\subfloat[Fastest service]{\includegraphics[width=0.48\textwidth]{Figures/App_Distribution_Fastest.pdf}
\label{fig:FastestApps}}
\hfil
\subfloat[ELECTRE]{\includegraphics[width=0.48\textwidth]{Figures/App_Distribution_ELECTRE.pdf}
\label{fig:ELECTREApps}}
\caption{The distribution of workloads between Fog nodes in the generic architecture.}
\label{fig:AppDist}
\end{figure*}

The implementations of the random offloading and the DRR selection algorithms are straightforward, where in the former, IoT devices randomly select one of the three Fog nodes to serve Sensor workloads. For DRR, the three Fog nodes are alternatively selected by each IoT device in a Round-Robin manner. With the nearest node selection algorithm, IoT devices select directly connected Fog nodes to serve Sensor workloads, which means that $Fog_3$ is never selected using this method. $Fog_2$ receives around 91\% of load as it is directly connected to around 91\% of IoT devices, while $Fog_1$ receives workloads from only two IoT devices. This approach, which is common in practice in Edge Computing environments, wastes compute and communication resources that shall be utilized to increase the overall system performance.

Using the fastest service selection algorithm, the Fog node with the smallest total execution delay is always selected, which is calculated by adding the service time to the network latency as we have discussed in the previous section. In the simple generic architecture, the fastest service for IoT devices 4 \& 5 is always their directly connected Fog node ($Fog_1$) (see Fig. \ref{fig:AppDist}\subref{fig:FastestApps}). However, for the rest of IoT devices, only less demanding workloads are sent to the directly connected Fog node, i.e., $Fog_2$, because of its scarce resources. While high-demanding workloads, i.e., from $App_3$, are sent to $Fog_3$ as it has the fastest service for such workload requirements. Our proposed approach, however, has a different behavior for IoT devices that are directly connected to $Fog_2$ (see Fig. \ref{fig:AppDist}\subref{fig:ELECTREApps}).

\begin{figure}[!t]
    \centering
    \includegraphics[width=0.48\textwidth]{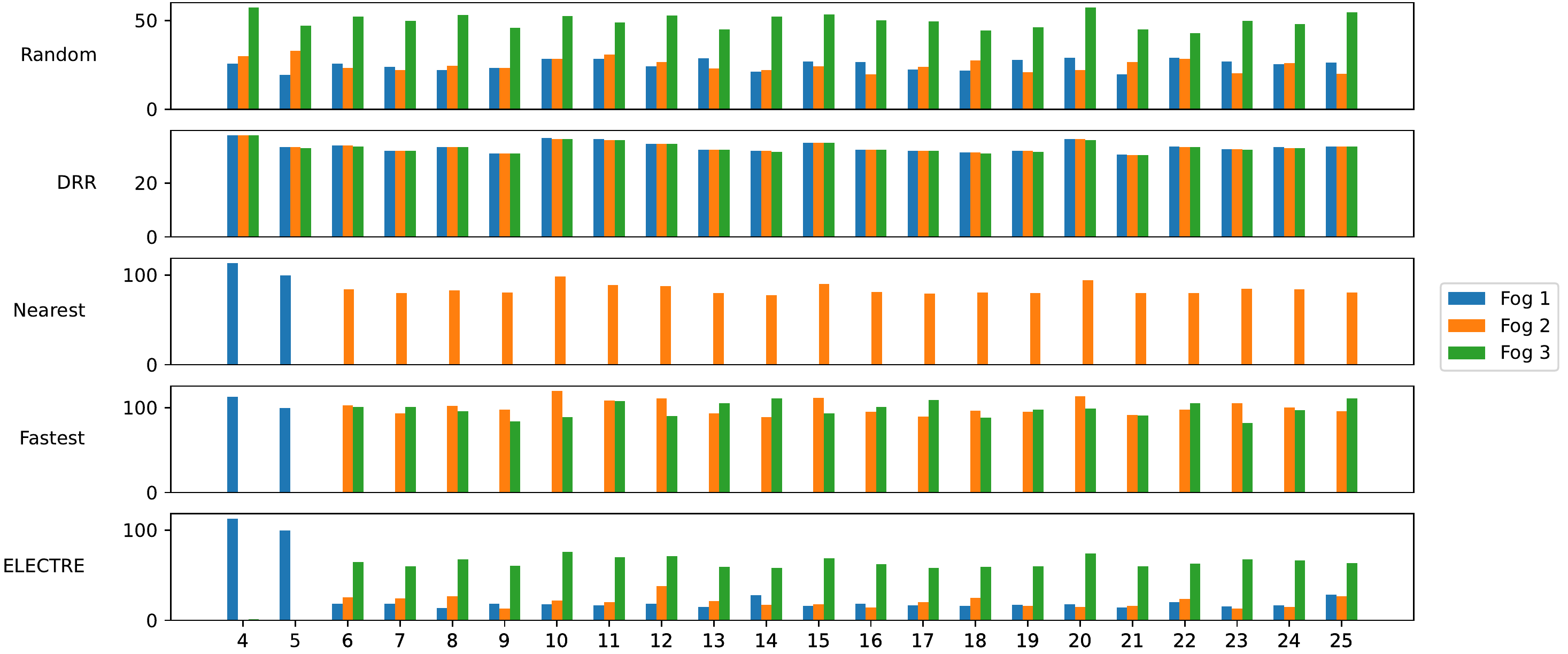}
    \caption{Number of received messages from each Sensor (over $10^4$ time-steps).}
    \label{fig:trans}
\end{figure}

ELECTRE distributes high-demand workloads, i.e., $App_3$ requests, between $Fog_1$ and $Fog_2$ to avoid the communication bottleneck between $Fog_3$ and the Cloud. This is because of the transmission of large workload data over limited network resources, which significantly affects the performance of the application loop as a whole. However, ELECTRE chooses to send smaller workloads to the fastest Fog node as they have less impact on network transmission between $Fog_3$ and the Cloud. Fig. \ref{fig:trans} confirms these findings, where $Fog_1$ is always selected for IoT devices 4 \& 5 using the nearest node, the fastest service, and the ELECTRE algorithms. While for the rest of IoT devices, $Fog_2$, $Fog_2$ \& $Fog_3$, and the three Fog nodes are chosen by these three algorithms, respectively. 

\begin{figure}[!t]
    \centering
    \includegraphics[width=0.30\textwidth]{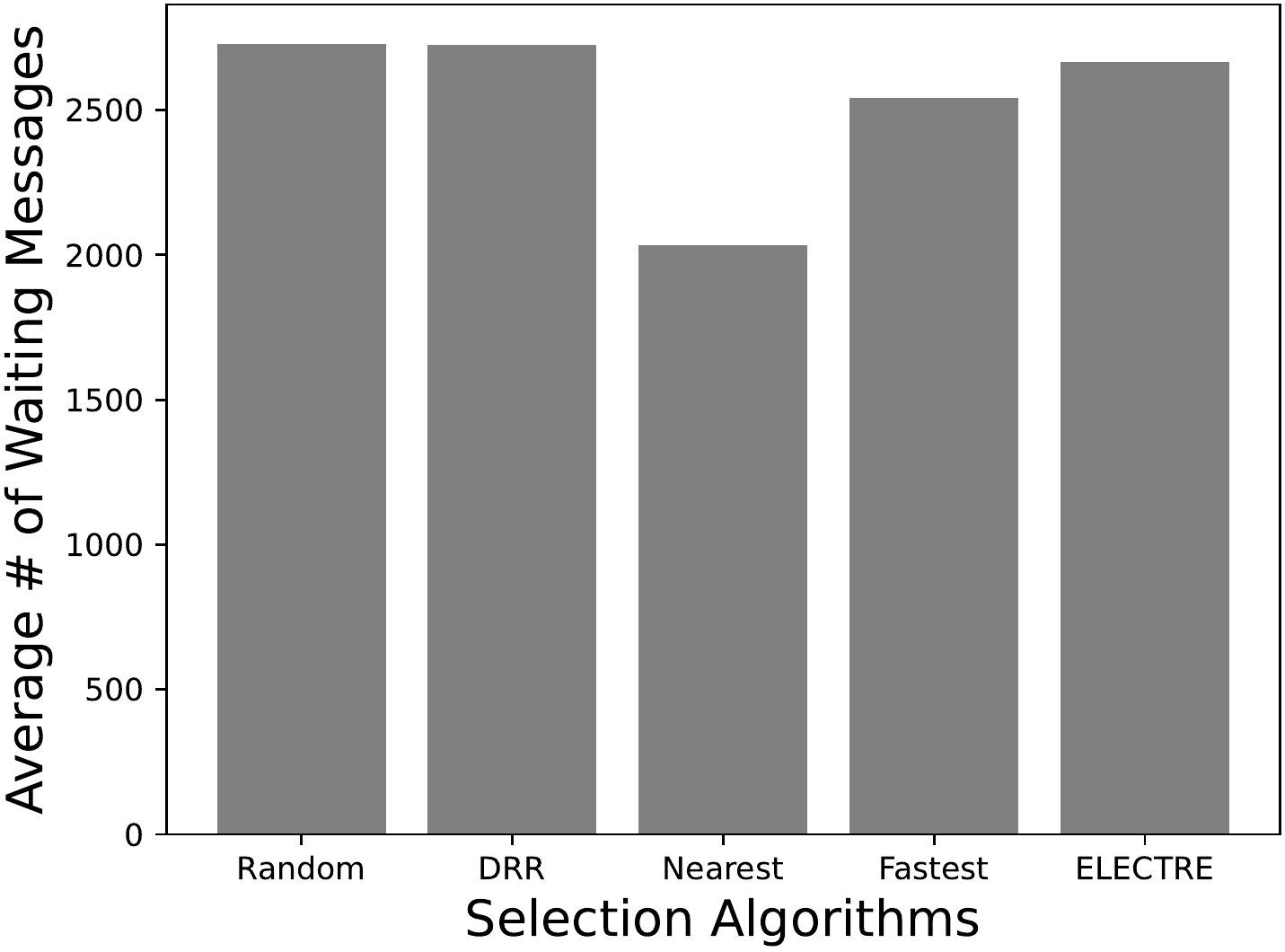}
    \caption{Network saturation (over the simulation duration of $10^4$ time-steps).}
    \label{fig:satu}
\end{figure}

Since the network is modeled as an M/M/1 queue model, we can calculate the number of messages that are, by the end of the simulation, still waiting to access a network link. Typically, these messages will have to stay in a network buffer during the simulation until they get their turn to access the network link, which is done in a First-Come-First-Served (FCFS) manner. Fig. \ref{fig:satu} shows that the nearest node selection algorithm produces the fewest number of waiting messages as it pumps the least number of messages into the network. The reason for this is choosing the slowest Fog node ($Fog_2$) 91\% of the time, because it is directly connected to the majority of IoT devices in our unbalanced generic architecture. Hence, $Fog_2$ will act as a computation bottleneck, emitting a smaller number of messages through the application workflow.

\begin{figure}[!t]
    \centering
    \includegraphics[width=0.48\textwidth]{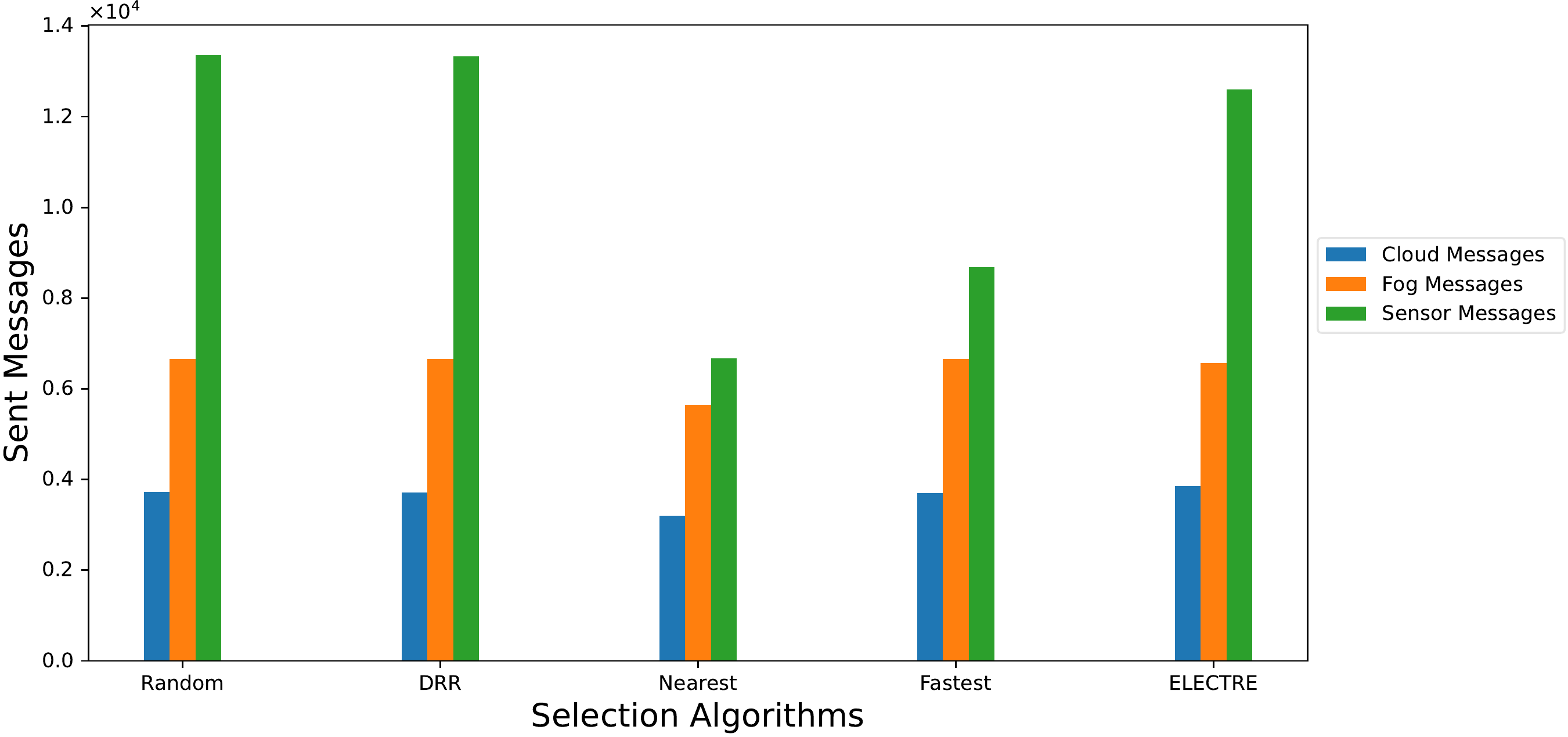}
    \caption{Number of transmitted messages (over $10^4$ time-steps).}
    \label{fig:sent}
\end{figure}

\begin{figure}[!t]
    \centering
    \includegraphics[width=0.48\textwidth]{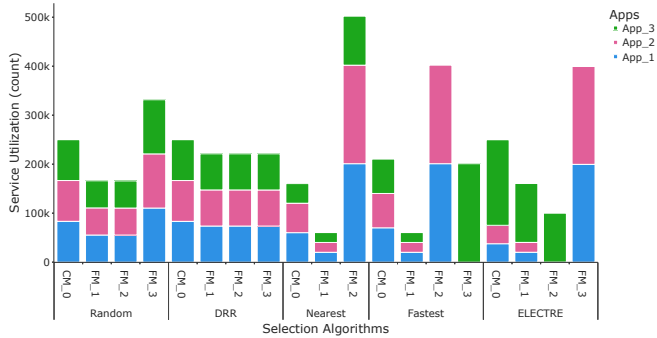}
    \caption{Number of served messages (over $10^6$ time-steps).}
    \label{fig:serve}
\end{figure}

Fig. \ref{fig:sent} supports this analogy by showing the number of transmitted messages, i.e., messages that traveled a communication link. It confirms that the nearest node selection algorithm sends the least number of messages to the system for the reason discussed above. Fig. \ref{fig:serve} shows the number of messages that were served by each service, which also supports the same fact regarding $Fog_2$ acting as the bottleneck of the system due to its poor computation resources. The number of workloads served by the Cloud module using the nearest service selection algorithm is less than all other algorithms, which is due to the number of workloads served by the system computation bottleneck ($Fog_2$).

Our proposed ELECTRE algorithm smartly distribute workloads based on the workload's resource requirements and the current load of Fog nodes. This allows more workloads to be served at the Cloud, especially high-demanding workloads, i.e., from $App_3$. The reason for this is the intelligent workload assignment to the resource-rich Fog node, i.e., $Fog_3$ (see Fig. \ref{fig:AppDist}\subref{fig:ELECTREApps} \& Fig. \ref{fig:serve}). ELECTRE selects $Fog_3$ for workloads from $App_1$ \& $App_2$, and distributes resource-demanding workloads (from $App_3$) between $Fog_1$ \& $Fog_2$.

\begin{figure}[!t]
    \centering
    \includegraphics[width=0.4\textwidth]{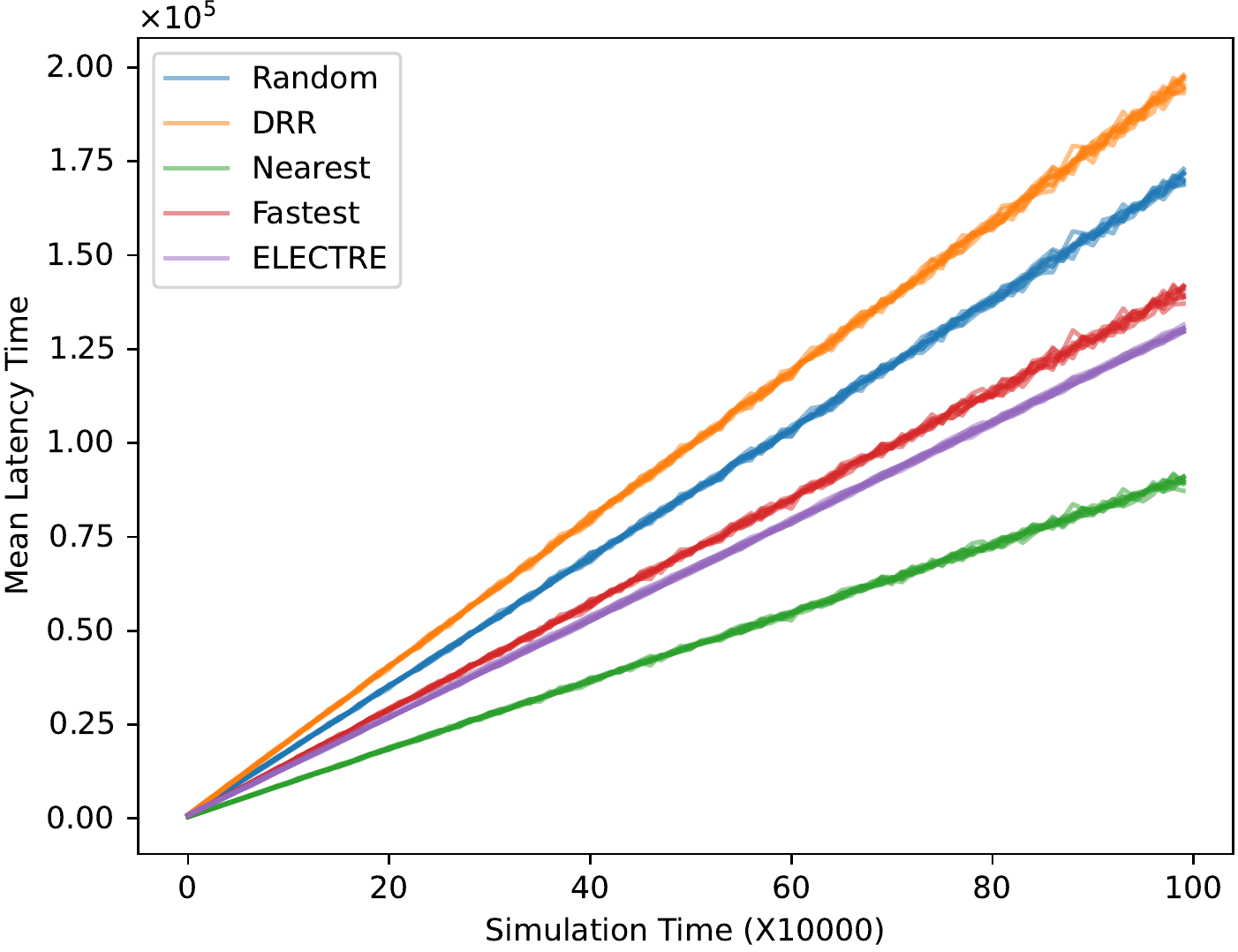}
    \caption{Network latency of transmitted messages (resampled every $10^4$ time-steps over the simulation duration of $10^6$ time-steps).}
    \label{fig:lat}
\end{figure}

In terms of network latency, Fig. \ref{fig:lat} shows how the nearest node selection algorithm produces the minimum mean latency values compared to the rest of the algorithms. However, this is a bit optimistic since it is calculated for all transmitted messages in the network. Hence, latency values for this algorithm must be smaller as it sends the smallest number of messages over the network compared to all other algorithms (see Fig. \ref{fig:satu}). Moreover, spending more time processing workloads in the system's bottleneck ($Fog_2$) causes a small number of Fog messages to be sent over network links, and hence a small number of Cloud messages will be sent accordingly. Because a small number of messages compete for links' resources using this algorithm, the waiting time for these messages to access network links will be smaller.

In addition, Sensor messages only travel a single link from source to destination using the nearest node selection algorithm. Unlike the rest of the algorithms, where they might travel multiple links to reach their destination. This adds to why latency values are smaller than other algorithms using the nearest node selection algorithm. ELECTRE achieved the second-best performance in terms of latency even with Sensor messages traveling multiple links to reach their destination Fog node. However, latency here is more realistic as more messages are generated using ELECTRE, compared to the nearest node selection algorithm (see Fig. \ref{fig:satu}).

\begin{figure}[!t]
    \centering
    \includegraphics[width=0.48\textwidth]{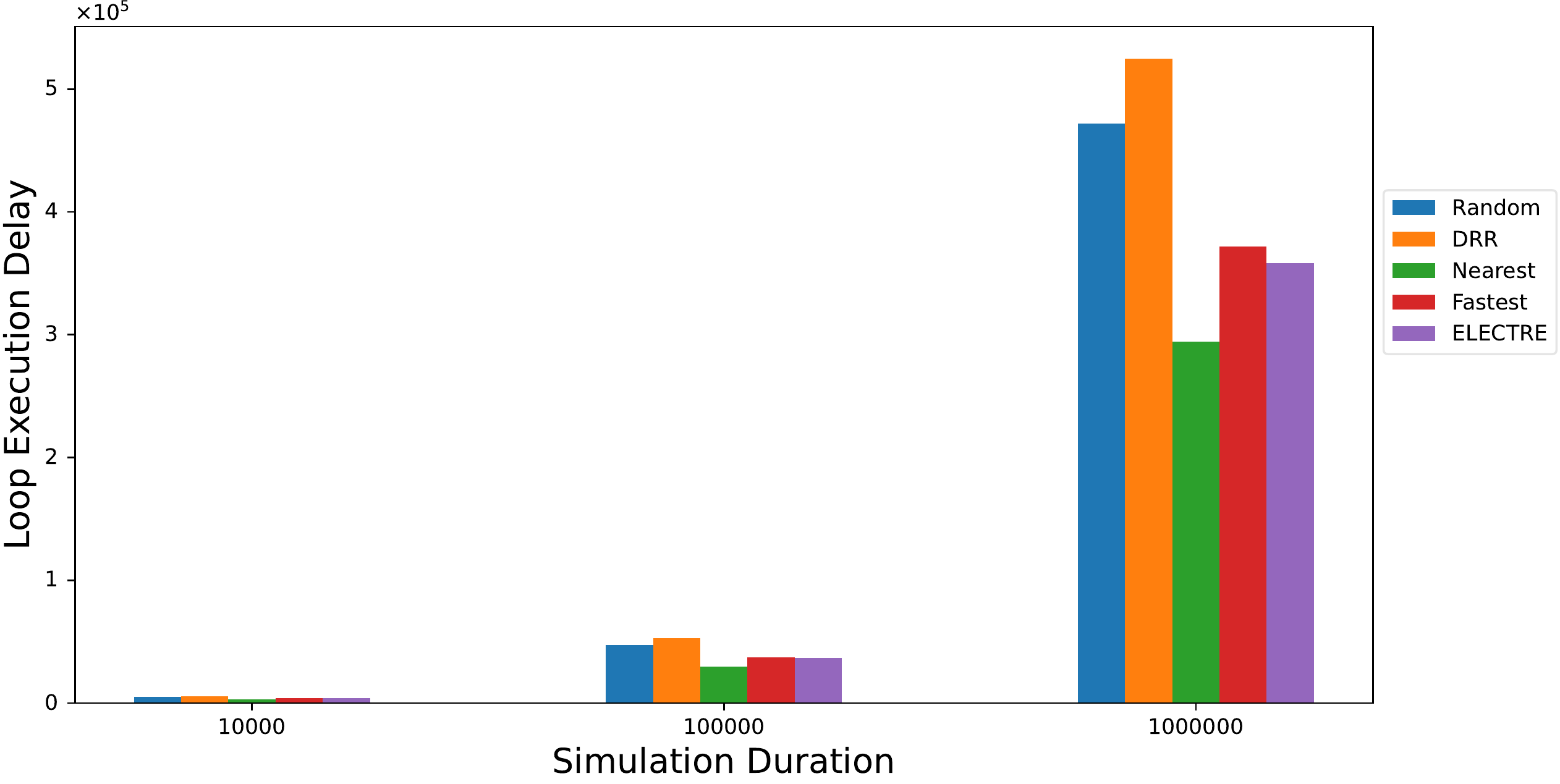}
    \caption{Mean Execution delay for the three distributed application loops.}
    \label{fig:lop}
\end{figure}

After evaluating the performance of these algorithms in terms of individual messages, we now evaluate their performance in terms of the whole application loop by accumulating the execution delay of each message in that loop (see Fig. \ref{fig:lop}). The nearest node selection algorithm achieved the minimum mean loop execution delay. This is too optimistic, as stated earlier, because this method sends the least number of messages along each application loop. In contrast, ELECTRE achieved the second-best mean loop execution delay while sending more messages along that loop. To provide a fair comparison between the performance of each algorithm in our generic architecture, we evaluate the loop execution delay in terms of the number of messages, or bytes, transmitted along that loop.

\begin{figure}[!t]
    \centering
    \includegraphics[width=0.48\textwidth]{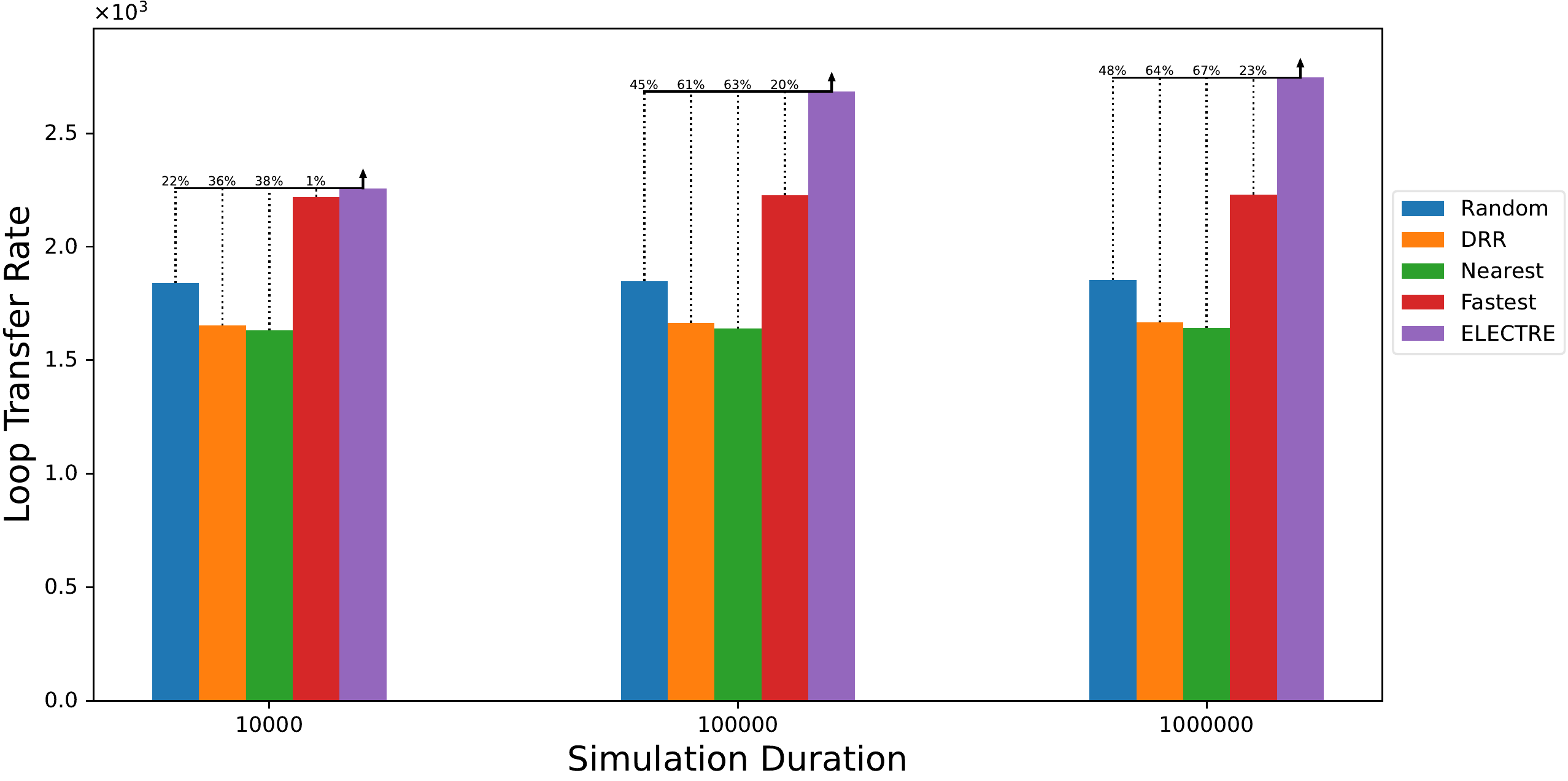}
    \caption{Mean Loop Transfer Rate (overall system performance).}
    \label{fig:op}
\end{figure}

This is done using what we call the mean loop transfer rate (see Fig. \ref{fig:op}), which is calculated by dividing the number of transmitted bytes along an application loop by the mean execution delay of that loop. The mean loop transfer rate, measured in Bytes per Second (BPS), can be viewed as the overall performance of that loop. This performance measure provides fairness by considering a good algorithm to be the one that transmits as many messages as possible while maintaining a small loop execution delay. Using this metric, in Fig. \ref{fig:op}, DRR and nearest node selection algorithms achieve the worst performance compared to the rest of the algorithms. The random-based approach achieves a slightly better performance by increasing the system's resource utilization using random load distribution between Fog nodes.

Our proposed ELECTRE-based method achieved the best overall system performance, i.e., highest loop transfer rate, compared to other methods in this study. Over $10^4$ simulation time-steps, our approach achieves a 1\% improvement over the second best approach in our study, i.e., fastest service selection. However, the performance of our approach significantly increases when running the simulation over longer time-steps, i.e., $10^5$ and $10^6$, where it achieves 20\% and 23\% improvement over the second best approach, respectively. Thus, over $10^6$ time-steps, our approach achieved an improvement between 23\% and 67\% over the other approaches used in this study.

\begin{figure}[!t]
    \centering
    \includegraphics[width=0.48\textwidth]{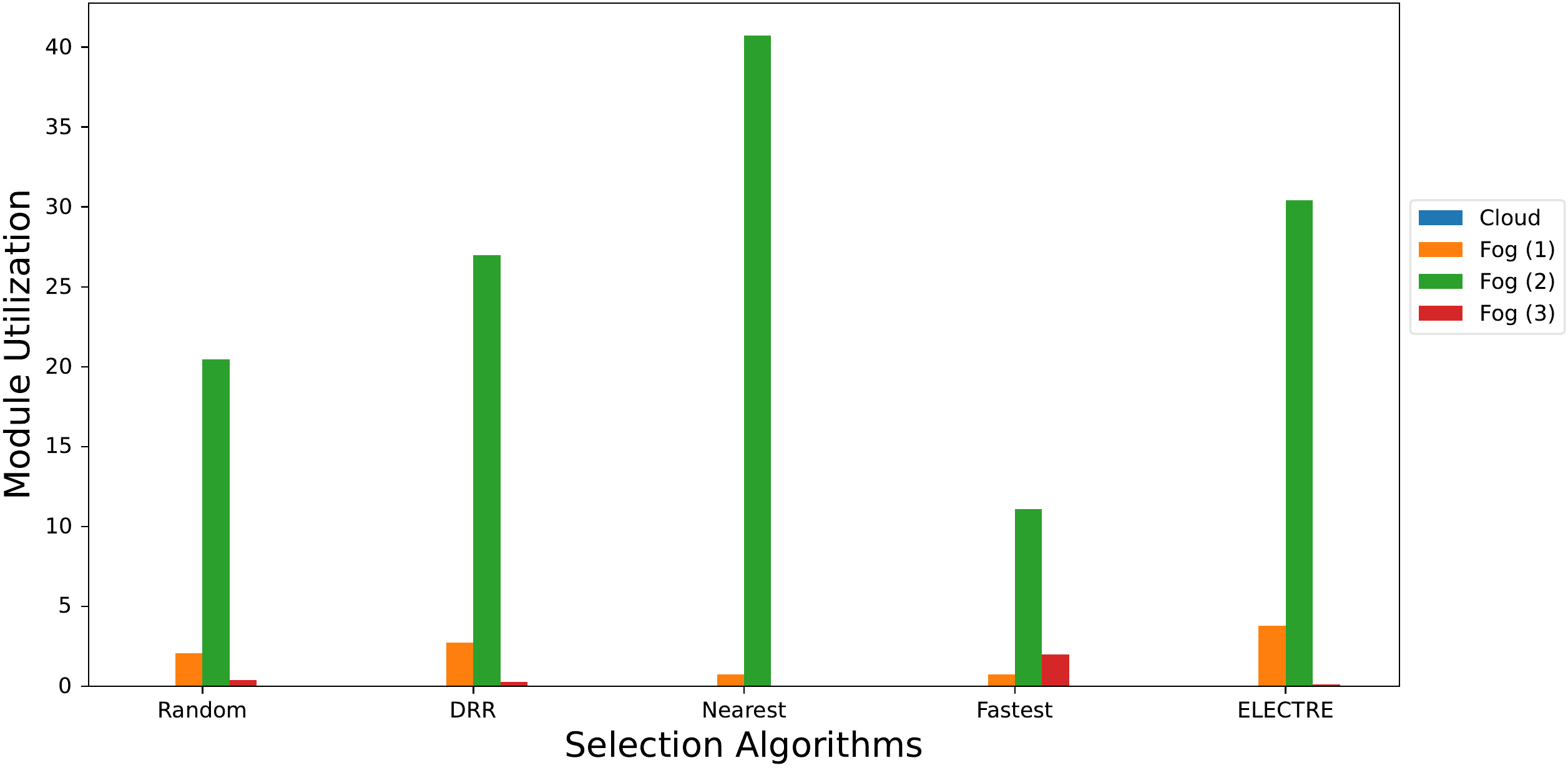}
    \caption{Module Utilization for each message type (over $10^4$ time-steps).}
    \label{fig:util}
\end{figure}

Fig. \ref{fig:util} shows the module utilization that was achieved by each one of these algorithms. The fastest Fog node ($Fog_3$) is never utilized using the nearest node selection algorithm, while it is used the most using the fastest service selection algorithm. In contrast, the slowest Fog node ($Fog_2$) was used the most using the nearest node selection algorithm, while it was used the least using the fastest service selection algorithm. A better utilization for both $Fog_1$ and $Fog_2$ is very important since they are directly connected to IoT devices. Unlike $Fog_3$, which must be carefully used since it requires traveling more network links to be reached from IoT devices. 

In our generic architecture, we see this behavior using ELECTRE, where it utilizes $Fog_3$ for lighter workloads as discussed earlier (see Fig. \ref{fig:AppDist}\subref{fig:ELECTREApps}). On the other hand, it distributes resource-demanding workloads that are initiated from the majority of IoT devices, i.e., those connected to the slowest Fog node, between $Fog_1$ \& $Fog_2$. This is why $Fog_1$ is utilized the most using ELECTRE, while it uses $Fog_2$ more than all other algorithms except the nearest node method. Considering workload requirements, resource capabilities of Fog nodes, and their resource availability allows our proposed method to outperform other traditional methods with better load balancing in this generic unbalanced Fog architecture.

\begin{figure}[!t]
\centering
\includegraphics[width=0.48\textwidth]{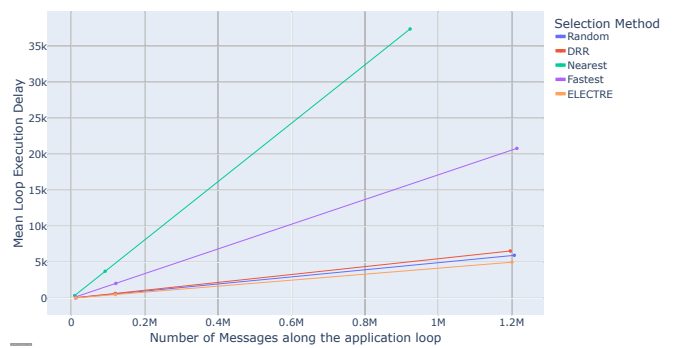}
\caption{The growth of the Mean Loop Execution delay and the number of served requests in the AS-inspired topology over three simulation durations.}
\label{fig:lop2}
\end{figure}

To confirm our findings in the generic architecture, we evaluate these algorithms in the Fog environment shown in Fig. \ref{fig:topo2} using the distributed application workflow shown in Fig. \ref{fig:2loops}. Figure \ref{fig:lop2} shows the performance of the five service selection algorithms in the AS-inspired topology averaged over the two application loops of each application. It shows how the mean loop execution delay and the number of served requests increase by increasing the simulation duration in our experiments. Figure \ref{fig:lop2} shows how the nearest node selection algorithm performs the worst in this environment, i.e., with the largest mean loop execution delay and the smallest number of requests served along the application loops.

\begin{figure*}[!t]
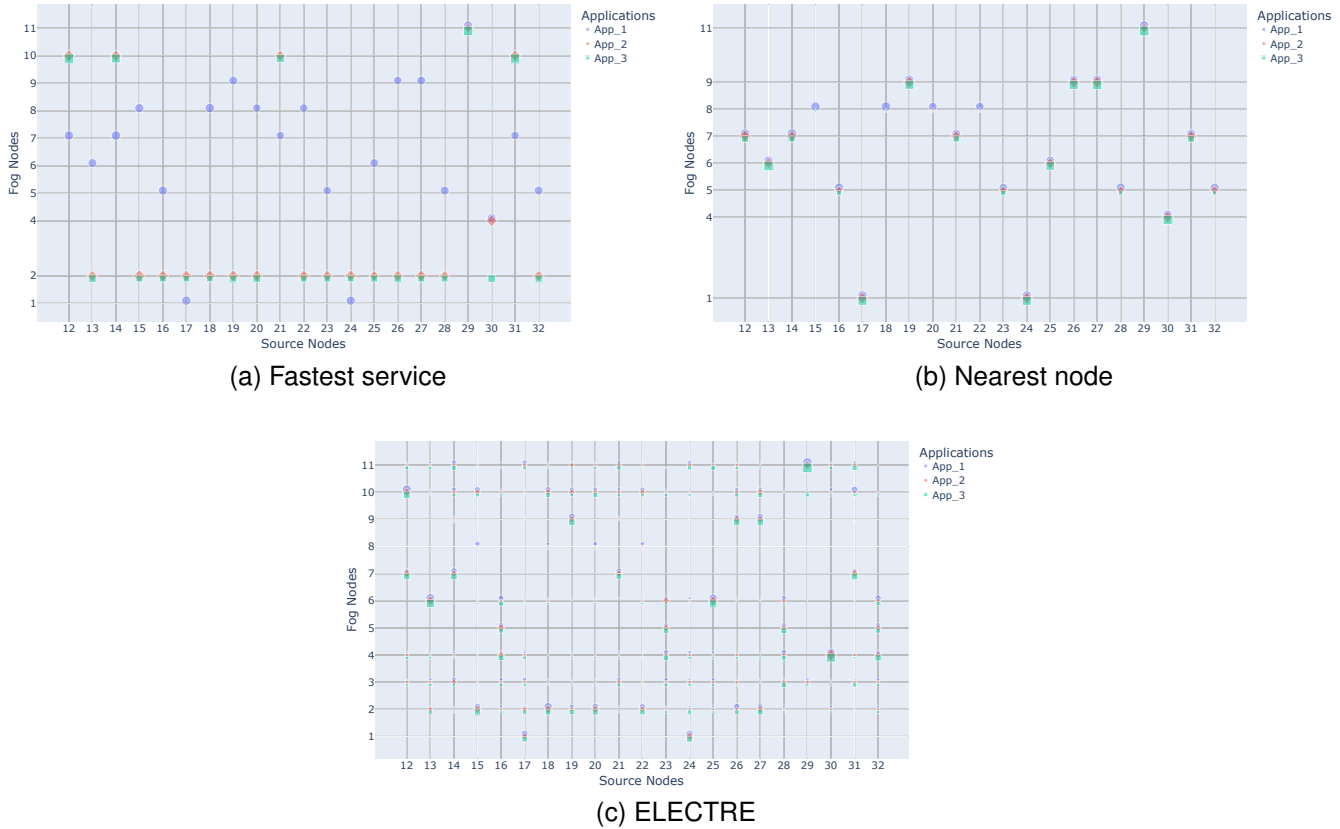

\centering
\subfloat[Fastest service]{\includegraphics[width=0.48\textwidth]{Figures/App_Distribution_Fastest_2.pdf}
\label{fig:FastestApps_2}}
\hfil
\subfloat[Nearest node]{\includegraphics[width=0.48\textwidth]{Figures/App_Distribution_Nearest_2.pdf}
\label{fig:NearestApps_2}}
\hfil
\subfloat[ELECTRE]{\includegraphics[width=0.48\textwidth]{Figures/App_Distribution_ELECTRE_2.pdf}
\label{fig:ELECTREApps_2}}
\caption{The distribution of workloads in the AS-inspired complex Fog topology.}
\label{fig:AppDist_2}
\end{figure*}

The fastest service selection algorithm has the second worst performance in terms of the mean loop execution delay. However, the largest number of served requests was achieved using this method. Although, the number of served requests along application loops using our proposed method is very closed to that achieved by the fastest service selection algorithm. Nevertheless, our proposed method achieved the best mean loop execution delay in this complex environment. To check why the nearest node and the fastest service selection algorithms performed the worst in the AS-inspired architecture, we examine the distribution of workloads in this environment (see Fig. \ref{fig:AppDist_2}).

In Figs. \ref{fig:AppDist_2}\subref{fig:FastestApps_2} \& \ref{fig:AppDist_2}\subref{fig:NearestApps_2}, we see the distribution of workloads from IoT devices to Fog nodes in the AS-inspired architecture using the fastest service and the nearest node selection algorithms, respectively. Fog nodes 2 \& 3 were never selected by any IoT device using the nearest node selection algorithm since they were not directly connected to any IoT device. Using the fastest service selection algorithm, Fog node 3 was never used while Fog node 2 is used for most resource-demanding workloads. The reason for choosing Fog node 2 for these type of workloads is having more resources compared to the resources of directly connected Fog nodes of most IoT devices (see Fig. \ref{fig:topo2}). Interestingly, our proposed ELECTRE-based method smartly distributes the load between all Fog nodes in the system (see Fig. \ref{fig:AppDist_2}\subref{fig:ELECTREApps_2}).

\begin{figure*}[!t]
\centering
\subfloat[Latency]{\includegraphics[width=0.48\textwidth]{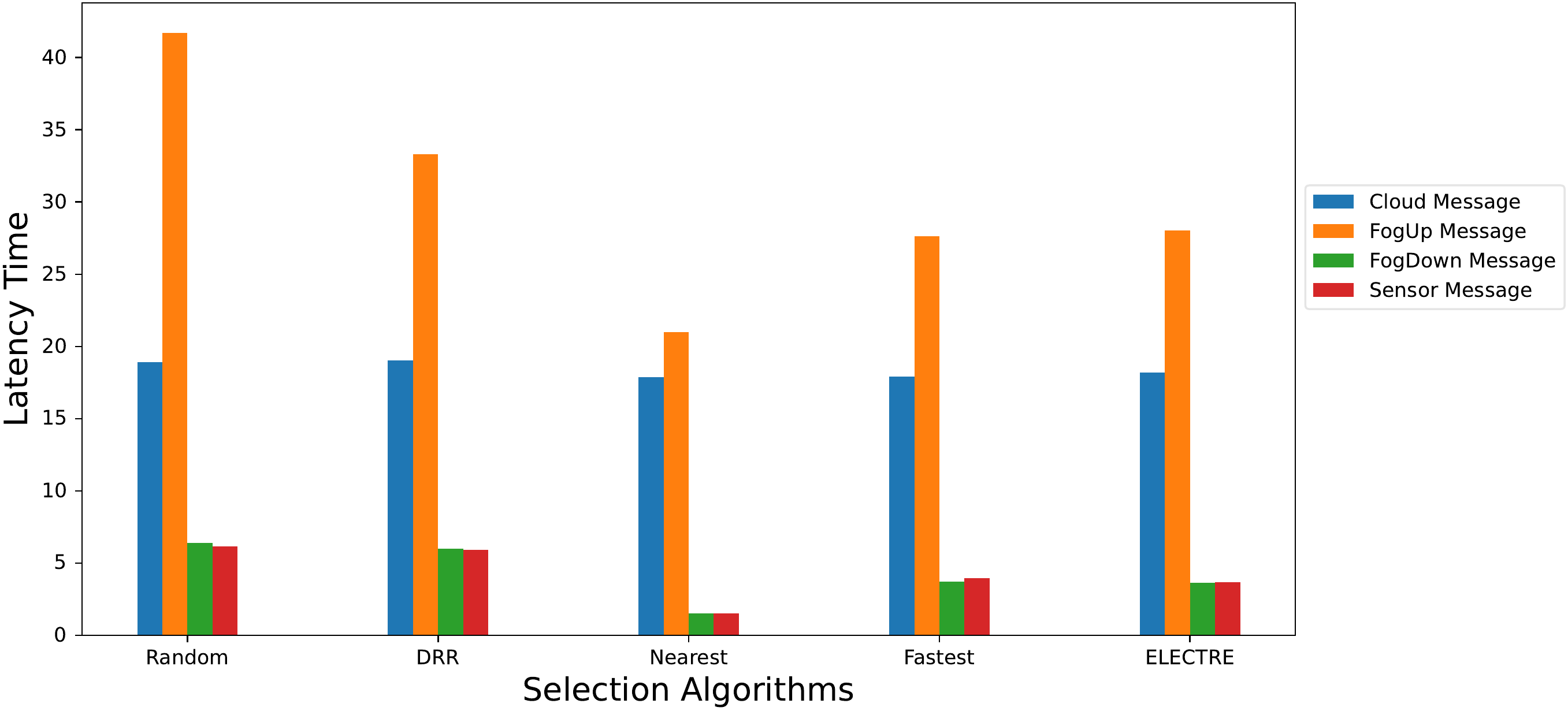}
\label{fig:time_latency}}
\hfil
\subfloat[Waiting Delay]{\includegraphics[width=0.48\textwidth]{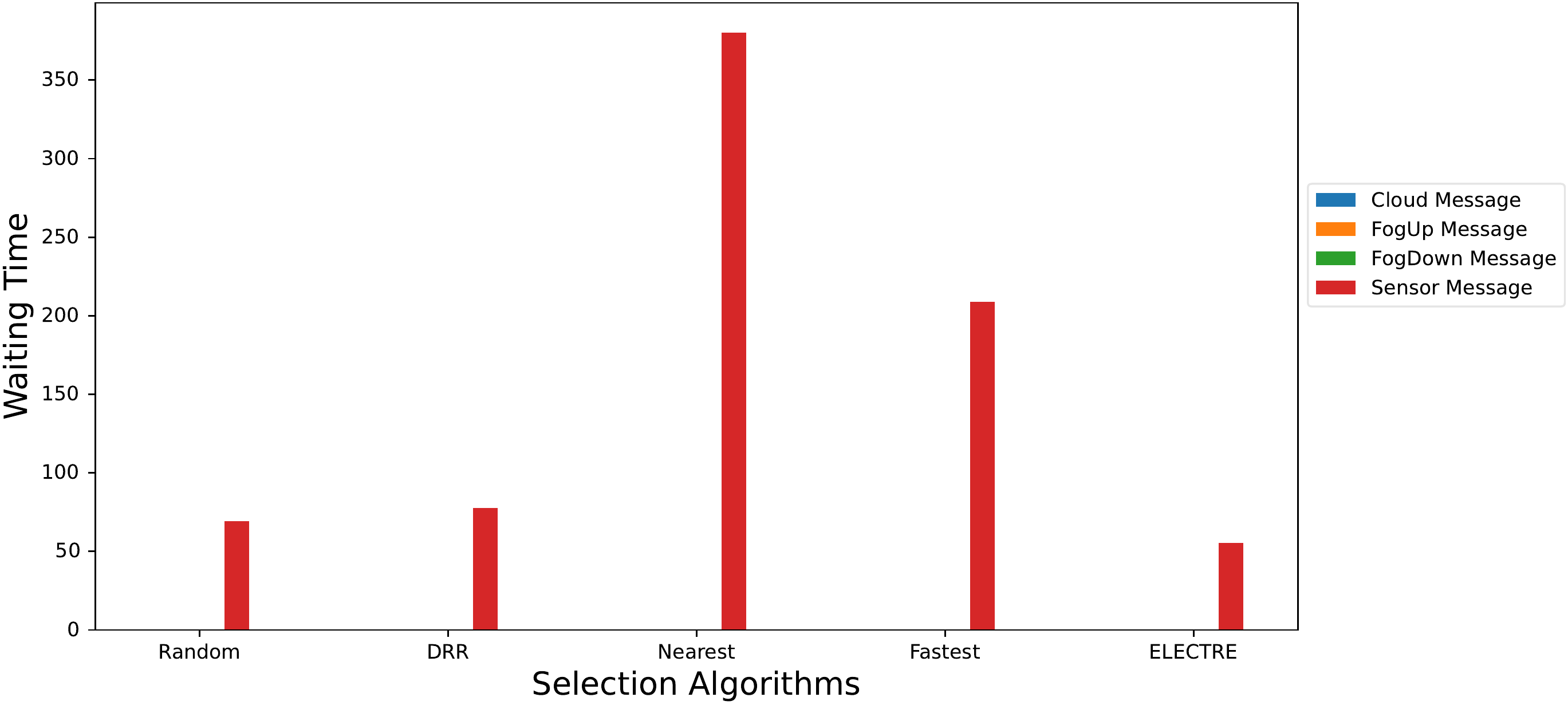}
\label{fig:time_waiting}}
\hfil
\subfloat[Service Time]{\includegraphics[width=0.48\textwidth]{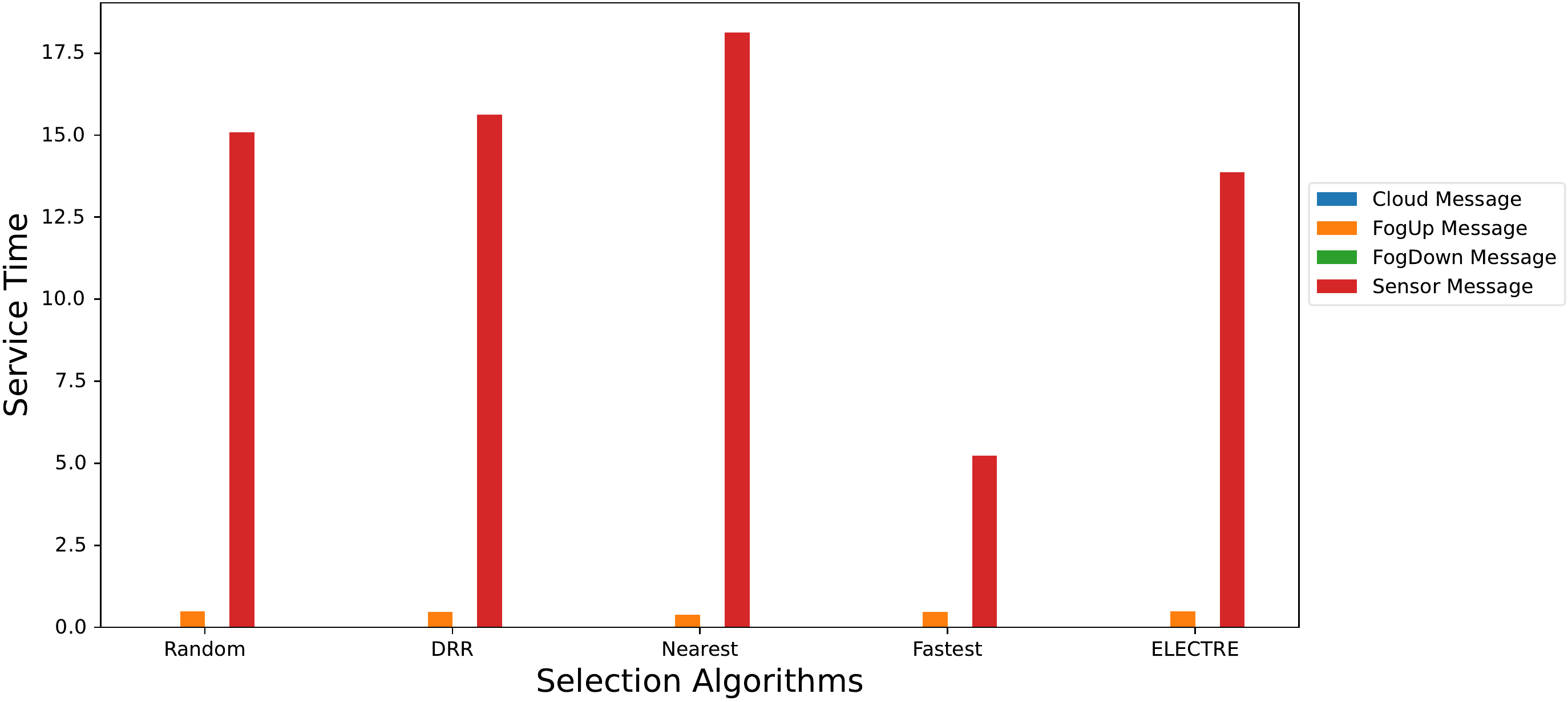}
\label{fig:time_service}}
\hfil
\subfloat[Response Time]{\includegraphics[width=0.48\textwidth]{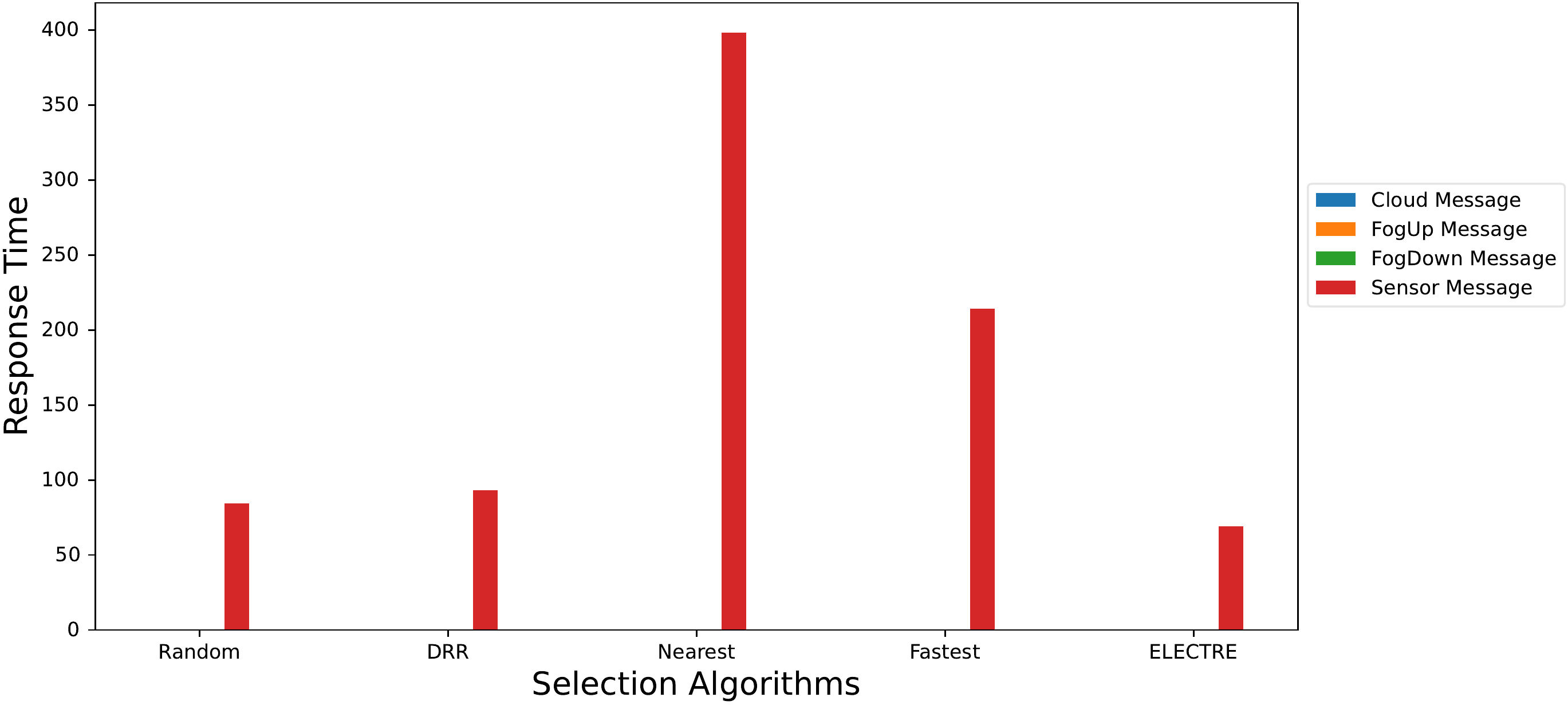}
\label{fig:time_response}}
\hfil
\subfloat[Total Response]{\includegraphics[width=0.48\textwidth]{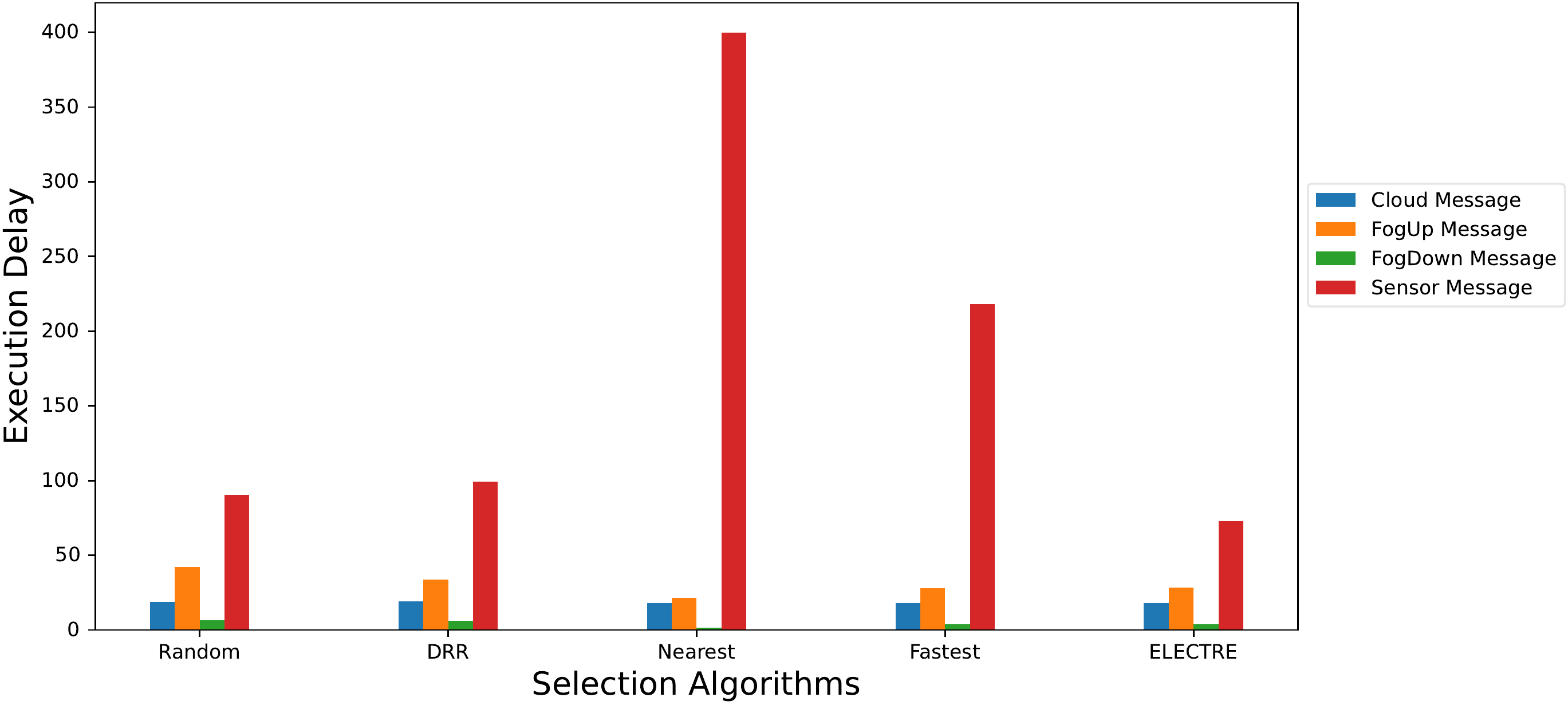}
\label{fig:time_total}}

\caption{Average delays in the AS-inspired complex Fog topology.}
\label{fig:Delays_2}
\end{figure*}

ELECTRE distributes the load of every application module between a number of Fog nodes, which increases their resource utilization while minimizing the number of waiting requests in each node. This allows the ELECTRE algorithm to achieve the smallest average waiting delay in the AS-inspired environment (see Fig. \ref{fig:Delays_2}\subref{fig:time_waiting}). Although, the nearest node and fastest service selection algorithms achieved the smallest average latency and service time (Figs. \ref{fig:Delays_2}\subref{fig:time_latency} and \ref{fig:Delays_2}\subref{fig:time_service}), respectively. However, the effect of reducing the waiting delay is significant, which is why ELECTRE has achieved the best response time (see Fig. \ref{fig:Delays_2}\subref{fig:time_response}), and the best execution delay accordingly (total response time in Fig. \ref{fig:Delays_2}\subref{fig:time_total}).

The results on the randomized AS-inspired Fog architectures confirm our findings on the generic Fog architecture. These results help understand the reasons for outperforming traditional baseline approaches using our proposed method. However, this performance gain does not come without a cost, which is in our case the computational complexity of the ELECTRE algorithm compared to these simple traditional methods. Still, this complexity can be acceptable when the performance gain is significant for the whole system. Moreover, the algorithm can run in resource-rich controllers that act as load balancers for the system to mitigate its complexity. This can be done through SDN-based Fog systems, where SDN controllers make offloading decisions to distribute the load in the system. In addition, ELECTRE can be used as a baseline to evaluate new load balancing algorithms, besides traditional baseline approaches.

\section{Conclusion}
\label{sec:conclusion}
Fog Computing is one of the main building blocks that enable the development of distributed, real-time, and delay-sensitive IoT applications. IoT devices are often resource-limited and battery-powered, which calls for the exploitation of Internet-based computing and storage resources. The Cloud was the first source of such resources, but Fog Computing quickly emerged to provide these resources in proximity to IoT devices. However, the vast number of IoT devices requires efficient utilization of such distributed resources, which can be only done through efficient load distribution. But load balancing in the Fog is very complex due to various factors, including but not limited to:
\begin{itemize}
    \item Heterogeneity of Fog, IoT, and network resources as well as IoT application modules.
    \item Using undedicated, public, and slow network links.
    \item Fluctuations in workload generation rates.
    \item Physical distribution of IoT devices and Fog nodes, i.e., IoT and Fog communities.
    \item Proximity of Fog resources to IoT devices, i.e., number of network links and communication latency.
    \item Fog/IoT mobility and their re-association process, i.e., dynamically changing environments.
\end{itemize}

In this work, we provide an efficient load balancing algorithm for Fog environments through efficient task assignment decisions, i.e., by selecting optimal Fog service replicas to serve IoT workloads. Considering stateless micro Fog service replicas provides network resilience in such environments by maintaining service availability in the face of faults and challenges to normal operation. To demonstrate the effectiveness of our approach in a realistic setup, we evaluated our approach in two Fog architectures with unbalanced resource and load distribution, heterogeneous resources, heterogeneous workload requirements, and a semi-hierarchical topology.

We start evaluating our approach on a simple and generic Fog architecture, which can be used to evaluate, evolve, and advance load balancing algorithms. Then, we evaluate our proposed solution in a randomized, complex, and realistic Fog setup that is inspired from the architecture of Internet Autonomous System (AS) networks. We compared our ELECTRE-based MCDA method with four traditional service selection algorithms that are commonly used in practice. The first one is a random-based algorithm inspired from a sequential randomization load balancing solution from the literature. This simple randomized approach exploits the power of random choice property to achieve good results with minimal overhead. In addition, we also implemented DRR, nearest node, and fastest service selection algorithms to be compared against our proposed approach. 

The results of this study helps understanding the importance of efficient load distribution in unbalanced Fog networks, especially with resource-demanding heterogeneous workload requirements. Efficient load distribution is achieved by increasing Fog resource utilization while minimizing the average loop execution delay of distributed IoT applications. Our proposed ELECTRE-based method outperformed the other traditional baseline methods used in this study with improvements of up to 67\% in the generic architecture. Our proposed method has also outperformed the other methods in the AS-inspired randomized architecture over 10 randomized experiment trials.

Using our proposed approach, more messages were sent over the network while achieving a better loop transfer rate, mean execution delay, mean network latency, and utilization of computing resources in Fog nodes. However, this performance gain comes with the cost of requiring more storage and processing power when compared to simple traditional methods. However, this overhead can be acceptable when there is a significant gain for the overall system performance. In addition, the overhead of running these algorithms can be minimized when load balancing is done through resource-rich network controllers, like SDN Controllers.

As a future work, we want to evaluate approximation-based RL algorithms in large-scale Fog environments. The use of Deep Neural Networks (DNN) in approximation-based solutions makes them suitable to run in resource-limited devices, like IoT, Edge, and Fog devices. In addition, RL agents can learn in partially observable environments, which will allow them to perform without the need for collecting real-time load information from every computing node in the system. Instead, approximating load information can be enough to provide efficient load distribution between Fog nodes while minimizing the overhead of transmitting real-time load information over the network. Hence, using such lightweight solutions will allow implementing efficient distributed load balancing algorithms that minimize compute, storage, and network overhead.

\bibliographystyle{IEEEtran}
\bibliography{main.bib}
\newpage

\begin{IEEEbiography}[{\includegraphics[width=1in,height=1.25in,clip,keepaspectratio]{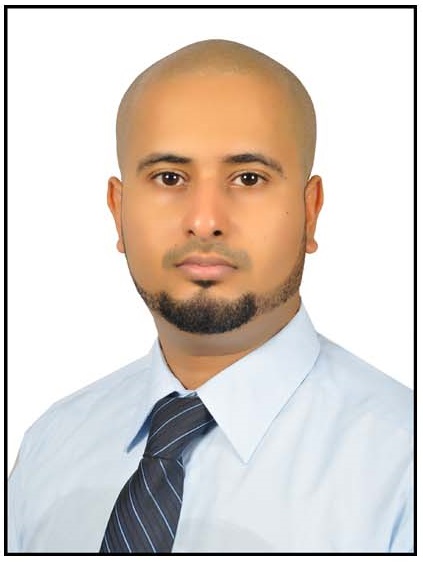}}]{Maad Ebrahim} is currently a Ph.D. candidate at the Department of Computer Science and Operations Research (DIRO), University of Montreal, Canada. He received his M.Sc. degree in 2019 from the Computer Science Department, Faculty of Computer and Information Technology, Jordan University of Science and Technology, Jordan. His B.Sc. degree in Computer Science and Engineering has been received from the University of Aden, Yemen, in 2013. His research experience includes Computer Vision, Artificial Intelligence, Machine learning, Deep Learning, Data Mining, and Data Analysis. His current research interests include Fog and Edge Computing technologies, Internet of Things, Reinforcement Learning, and Blockchains.
\end{IEEEbiography}

\begin{IEEEbiography}[{\includegraphics[width=1in,height=1.25in,clip,keepaspectratio]{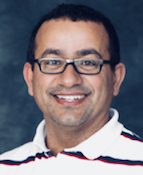}}]{Abdelhakim Hafid}
spent several years as the Senior Research Scientist with Bell Communications Research (Bellcore), NJ, USA, working in the context of major research projects on the management of next generation networks. He was also an Assistant Professor with Western University (WU), Canada, the Research Director of Advance Communication Engineering Center (venture established by WU, Bell Canada, and Bay Networks), Canada, a Researcher with CRIM, Canada, the Visiting Scientist with GMD-Fokus, Germany, and a Visiting Professor with the University of Evry, France. He is currently a Full Professor with the University of Montreal. He is also the Founding Director of the Network Research Laboratory and Montreal Blockchain Laboratory. He is a Research Fellow with CIRRELT, Montreal, Canada. He has extensive academic and industrial research experience in the area of the management and design of next generation networks. His current research interests include the IoT, fog/edge computing, blockchain, and intelligent transport systems.
\end{IEEEbiography}

\vfill
\end{document}